\documentclass[11pt,a4paper]{article}
\usepackage[utf8x]{inputenc}
\usepackage{ucs}
\usepackage[left=2.5cm,top=3cm,right=2cm,bottom=3cm]{geometry}
\usepackage{authblk}
\usepackage[authoryear]{natbib}
\usepackage{amsmath,amsfonts,bm,amssymb,latexsym,amsthm}
\usepackage{graphicx, graphics}
\usepackage{multirow}
\usepackage[authoryear]{natbib}
\usepackage{microtype}
\usepackage[none]{hyphenat}
\usepackage{setspace} 
\usepackage{array}
\usepackage{amsfonts}   
\usepackage[bottom]{footmisc}
\usepackage{color}
\usepackage{xurl}
\usepackage{float}
\usepackage{rotating}

\usepackage{soul} 
\usepackage{todonotes} 

\usepackage{placeins}
\usepackage{multicol}
\usepackage{multirow}

\usepackage{morefloats}
\usepackage{booktabs}
\usepackage{multirow}
\usepackage{pifont}

\usepackage{svg}

\usepackage{comment}

\newcolumntype{C}[1]{>{\centering\let\newline\\\arraybackslash\hspace{0pt}}m{#1}}

\usepackage{hyperref}
\hypersetup{colorlinks, breaklinks, linkcolor=[rgb]{0,0.3,1}, citecolor=[rgb]{0,0.3,1}, urlcolor=[rgb]{0,0.3,1} }
\usepackage{longtable}

\usepackage{lmodern} 

\begin{document}

\title{Are EU low-carbon structural funds efficient in reducing emissions?}

\author[1,3,*]{Marco Due\~nas}
\author[2,3]{Antoine Mandel}
\affil[1]{\small Institute of Economics -- Sant'Anna School of Advanced Studies, Pisa, Italy}
\affil[2]{\small Centre d'Economie de la Sorbonne --  Paris School of Economics -- CNRS-Universit\'{e} Paris 1 Panth\'{e}on-Sorbonne, Paris, France}
\affil[3]{\small CLIMAFIN -- Climate Finance Alpha, Paris, France}
\affil[*]{Corresponding author email: marco.duenas@climafin.com}

\date{}

\maketitle
\vspace{-40pt}

\begin{abstract} 
\noindent 
We investigate the effectiveness of low-carbon expenditures from the European Structural and Investment Funds in reducing greenhouse gas emissions across EU regions. Using trend and cycle decomposition of per capita emissions and emissions intensity, along with a panel data approach that incorporates long lags to mitigate reverse causality, we find highly heterogeneous effects. In less developed regions, investments are associated with long-term increases in per capita emissions, whereas in transition and developed regions, the effects are weak or not significant. When disaggregated by gas type, results remain inconsistent.  Our findings highlight that regional disparities challenge the effectiveness of EU climate efforts. 
\end{abstract}



\clearpage
\newpage

\section{Introduction}

The European Union (EU) has the objective of achieving climate neutrality by 2050. This implies large-scale investments in the energy transition \citep{mccollum2018energy, klaassen2023meta}. According to the European Commission, energy investments in the EU will have to reach €396 billion per year from 2021 to 2030 and €520-575 billion per year in the subsequent decades until 2050 \citep{energy_transition_in_the_eu}. A substantial share of these ought to be public investments that foster the development of infrastructure, the deployment of technological innovations, and the scale-up of private capital \citep{carraro2012investments,owen2018enabling,meckling2022busting}. Accordingly, the EU has enacted climate spending targets amounting to 20$\%$ of its 2014-2020 budget and 30$\%$ of its 2021-2027 budget that ought to be devoted to climate action\footnote{See \cite{eca_climate_spending_2022}, which also emphasises that climate spending over the 2014-2020 period has been over-estimated.}. Therefore, it is reasonable to question the efficiency of European climate policy to determine if these expenses effectively lead to reducing greenhouse gas (GHG) emissions. 

This paper exploits regional variations in European climate spending to address this issue. The main investment vehicles of the EU are the European Structural and Investment Funds (ESI funds), which accounted for €535 billion over the period 2014-2020 \citep{EU_ESI_Funds_2014_2020}. By construction, these funds are spent heterogeneously. EU regions are grouped into three categories based on their relative development level (least developed regions, transition regions, and developed regions), and this determines their eligibility and their financing rate for structural funds. From the 2014-2020 programming period onwards, the usage of these funds has been tagged with a ``thematic objective''. In particular, the ``Low-Carbon Economy'' tag allows the identification of the funds that have been assigned to climate action. Although the programming period 2007-2013 did not define thematic objectives in the same way, the aims of the expenditures are also tagged by ``priority codes'', from which it is possible to determine the funds related to climate action consistently with the low-carbon economy thematic objective from 2014-2020.

Consequently, we use the total amount spent on the low-carbon economy objective for the 2007-2013 and 2014-2020 budgeting periods as a measure of public climate investment. We then employ an econometric panel data model to investigate the influence of these investments on GHG emissions in European regions between 2007 and 2022 at the NUTS 2 level. Our analysis centres on two emission indicators: per capita emissions and emissions intensity (relative to regional gross domestic product (GDP)). We focus on the relationship between economic growth and environmental sustainability, and, to that end, we decompose the emission time series of each region into trend and cycle components. This allows us to determine expenditures' short-term and long-term effects, considering that some regions have significantly advanced in terms of sustainability---achieving a given degree of decoupling---whereas many others have yet to do so. Our analysis also accounts for the direct impacts of policy by considering the OECD Environmental Policy Stringency (EPS) index \citep{EPS_oecd} and average regional political opinions as covariates. 

To address potential endogeneity and reverse causality, we rely on a design with long lags of all right-hand-side variables, different specifications with controls including region-specific industry-composition (proxied by sectoral Gross Value Added (GVA)), and region and year fixed effects. We find heterogeneous effects of low-carbon expenditures on GHG emissions across European regions. These impacts vary depending on the regional development level, investment period, and emission indicator. In the least developed regions, low-carbon expenditures are consistently associated with increases in long-term emissions per capita and, to a lesser extent, in emissions intensity. In contrast, results for developed and transition regions do not reveal significant reductions in long-term emissions trends, although there is some evidence of short-term mitigating effects, particularly for the 2007–2013 programme. 

When disaggregating emissions by gas type, we consistently find marked regional heterogeneity. The results reveal no significant reductions in emissions for developed regions, while in less developed regions, emissions---especially CO$_2$ and N$_2$O per capita---increase significantly. Effects in transition regions are mixed. Short-term dynamics also show variation by gas and region: investments under the 2007–2013 programme help reduce emissions volatility, particularly for CO$_2$ in less developed and developed regions, and for N$_2$O in transition regions.

Overall, our findings point to important challenges in the effectiveness and coherence of current EU climate investment strategies, particularly those guided by the ESI funds, in supporting long-term decarbonisation in less developed regions. However, rather than signalling a failure of climate investment itself, the uneven performance may reflect the persistent structural regional disparities, including differences in institutional capacity, infrastructure, and development trajectories, as suggested by the Kuznets environmental curve \citep{bataille2021industry, saraji2023challenges}. Investments' immediate impacts are not necessarily always aligned with long-term sustainability goals. For instance, it could be that investments below a certain level may increase energy consumption and raise GHG emissions. In many developing regions, investments are focused on establishing basic infrastructure to boost economic output and meet developmental needs. 

In addition, low-carbon economy investments might require substantial up-front costs (deploying heavy infrastructure), and sustainable energy systems demand upgrading grids, developing renewable energy sources, and improving energy resilience. Though essential in the long term, these investments can increase short-term emissions due to the intensive energy and resources required during their implementation phase. On another front, investments can be misused, for example, as a countercyclical buffer to stabilise the economy during recessions or in implementing large and inefficient infrastructures (aka ``white elephants''). 

While arguments behind the heterogeneous impacts vary across regions, they are certainly shaped by economic conditions, particularly the economic cycle. A recent strand of literature offers insights into the interplay between environmental impacts and the business cycle. \cite{doda2014evidence} finds that CO$_2$ emissions increase during economic upturns, exhibit greater volatility compared to GDP, and show a procyclicality that is more pronounced in higher GDP per capita economies, suggesting that wealthier economies have more consistent emission patterns through economic cycles. \cite{sheldon2017asymmetric} finds an asymmetrical response of emissions to the business cycle, which is attributed to the timing of capital investments and technological advancements. Moreover, \cite{klarl2020response} finds CO$_2$ emissions' elasticity to GDP changes to be significantly higher during recessions in the United States. These results demonstrate the critical role of investment decisions in shaping environmental impacts.

In addition, climate and energy policy tools have shown some degree of effectiveness, however, their success often relies on countries' stringency and the complementarity among strategies. \cite{diakoulaki2007decomposition} find that despite considerable efforts, the expected acceleration in decoupling post-Kyoto Protocol was not uniformly achieved. For Europe, \cite{naqvi2017fifty} and \cite{ivanova2017mapping} find significant disparities in decoupling across countries, sectors, pollutants, and regions. \cite{papiez2022does, papiez2021role} further report mixed results in the EU's long-term decoupling efforts, pointing to disparities between older and newer member states and emphasising the role of energy intensity and structure in emissions reductions. 

In light of these findings, our paper contributes to the understanding of decoupling economic growth from emissions by analysing how low-carbon investments from the ESI funds impact emissions reductions within the business cycle.  

The remainder of this study is structured as follows. Section~\ref{sec:literature} discusses the relevant literature and the conceptual framework of our study and situates our contribution. Section~\ref{sec:data} presents the data and methods. Section~\ref{sec:results} presents the empirical results of the econometric estimations. Section~\ref{sec:discussion} discusses the results and provides policy implications. Finally, Section~\ref{sec:conclusions} concludes.

\section{Decoupling emissions: related literature}
\label{sec:literature}

Our study builds on the extensive literature on the connection between economic growth and GHG emissions. Drawing from the  Environmental Kuznets Hypothesis, it is assumed that an increase in wealth generates greater demand from the population for a cleaner and healthier environment. This demand drives governments to implement stricter environmental regulations to protect the environment. In this way, while wealthier countries tend to have the resources and political will to adopt these policies, regions with more constrained economic resources may find it more challenging to establish high environmental standards \citep{oconnor1992managing, doyle2014eliminating}. 

Certainly, economic development can lead to improved environmental outcomes by decoupling emissions from economic output. This is central to sustainable development and indicates that an economy can grow without a proportional increase in environmental degradation. However, the process is influenced by a multitude of factors, not only including income levels but also other aspects such as consumption patterns and regional characteristics. \cite{shuai2017turning} analyse the country-by-country turning points of the carbon Kuznets curve, finding a positive correlation between a country's income level and its likelihood to fit the Kuznets hypothesis, with higher income economies achieving quicker tipping points than their poorer counterparts. While economic development level is linked to faster environmental improvement transitions, other variables such as consumption patterns and regional characteristics also play a relevant role. \cite{ivanova2017mapping} analyse household consumption across 177 regions in 27 EU countries, finding wide variations in per capita GHG emissions across regions (0.6 to 6.5 tCO$_2$e/cap in 2007). Regarding the drivers behind carbon footprints, the authors identified the relevance of household size, whether an area is urban or rural, the population's education level, expenditure patterns, local temperature, resource availability, and the carbon intensity of the local electricity mix. 

Yet a significant strand of literature doubts that the downward-sloping relationships between higher-income countries and pollution levels---which might imply a decoupling of economic activity with emissions---are enough to face climate change efficiently. Certainly, there are concerns about the possible shift in the production of pollution-intensive goods from developed to developing countries with less stringent environmental regulations \citep{hertwich2009carbon, levinson2023developed}. This would weaken the Kuznets hypothesis since developed regions may reduce their at-home pollution levels by offshoring the production of environmentally harmful goods to nations where environmental regulations are less strict \citep{fischer2013environmental}. While this points out that high-income countries should also look into reducing aggregate demand as a strategy for significant emissions reductions \citep{knight2014economic}, it also requires the implementation of agreements at different scales: regional and global \citep{oberthur1999kyoto}.

In this context, a growing body of literature indicates that international environmental agreements and climate policies are increasingly impacting emission trends, as reflected in the economic practices within countries, contributing to our understanding of the decoupling mechanism \citep{fischer2013environmental, hilden2014climate, haberl2020systematic}. Decoupling analyses aim to categorise the patterns of how emissions respond to economic growth---whether they increase at a slower rate, decrease, or stabilise as the economy grows (relative and absolute decoupling) or increase faster than the economy (coupling). 

\cite{cohen2018long} provide valuable quantitative analysis that helps recognise how major emitting countries are progressing in decoupling emissions. Using information from OECD countries for 2001-2015, \cite{chen2018decomposition} analyse how economic growth can be decoupled from CO$_2$ emissions, identifying three key forms: recessive, where emissions drop due to efficiency gains or energy mix shifts, not from economic growth; weak negative, where efficiency improvements do not fully offset emissions from economic expansion; and strong, where economic and population growth do not lead to higher emissions, thanks to effective policies, technological progress, and structural economic changes.

Building on this understanding of the decoupling patterns, it is imperative to analyse emissions in the context of business cycles due to their substantial implications for environmental policy and economic management, particularly regarding the relationship between economic activity, regulation, emissions and energy consumption \citep{fischer2013environmental}. In fact, emissions tend to be procyclical, increasing during economic booms and decreasing during recessions \citep{heutel2012should}. A fundamental question relates to the causal relationship between economic output and energy consumption, especially targeting business cycle fluctuations \citep{narayan2011energy}. \cite{thoma2004electrical} shows that changes in macroeconomic conditions cause significant changes in electrical usage, particularly in the commercial and industrial sectors. These energy consumption responses reinforce the procyclicality of emissions relative to GDP.

Therefore, the connection among economic growth, energy consumption, and GHG emissions is well established in the short term. For example, \cite{peters2012rapid} claim that economic downturns typically lead to a reduction in energy-intensive activities and lower CO$_2$ emissions. Notwithstanding, the authors discuss that the 2008-2009 financial crisis saw a quick dip in GDP followed by a swift rebound in emissions by 2010, driven by falling energy prices, government economic incentives, and prior economic growth in developing countries, quickly returning to high emission levels post-crisis. 

Similarly, \cite{doda2014evidence} finds that CO$_2$ emissions are procyclical and more pronounced in countries with higher GDP per capita. Hence, the procyclicality also implies that economic growth without specific interventions often leads to increased emissions. Another fact is that emissions exhibit greater volatility compared to GDP, responding more dramatically to economic changes, and the extent to which emissions volatility exceeds GDP volatility decreases with the income level of the economies \citep{doda2014evidence}. These patterns are also observed at the subnational level: \cite{duenas2025regional}, analysing NUTS 2 regions in Europe, further show that CO$_2$ emissions growth rates follow an asymmetric and fat-tailed distribution and, similarly, that emissions volatility scales inversely with the economic size of regions. This implies that more stable emission patterns through economic cycles in developed regions could be linked to more efficient energy use, a greater share of services in the economy, and potentially more effective environmental regulations and policies. In addition, emissions decrease more significantly during economic contractions than during expansions, suggesting an elastic response to economic downturns and inelasticity to economic growth \citep{sheldon2017asymmetric}.

Likewise, \cite{klarl2020response} examines the fluctuation of CO$_2$ emissions' elasticity to GDP in the United States from 1973 to 2015, revealing that this elasticity is not constant and varies with economic conditions. He showed that emissions respond more sensitively to GDP changes during recessions than periods of economic normality, with elasticity often exceeding one in downturns. This indicates that CO$_2$ emissions decrease more rapidly than economic output in recessions and also suggests that the capacity of an economy to influence environmental outcomes can be significantly conditioned by its current economic state.

Thus, the literature has established the relevance of economic policy in addressing both the cyclical sensitivity of emissions to economic activity and the long-term decoupling of emissions from economic growth. This is of particular interest in the context of the European Union, where studies have started to examine the role of policy in shaping both short-term dynamics and long-term environmental trends.

For example, \cite{naqvi2021decoupling} finds that, on average, EU regions have managed to decouple emissions from economic output, with the most significant emission reductions occurring before the 2008 financial crisis. After 2008, the trend towards decoupling weakened, with instances of re-coupling between emissions and economic output observed. In addition, the author finds that environmental policy effectiveness varies significantly across different types of emissions and is influenced by the income levels of the regions. Similarly, with the aim of monitoring the effectiveness of the EU's environmental policies, \cite{delgado2018cyclical} developed the EU-cyclical environmental performance index by applying dynamic factor analysis over the period 1950 to 2012. This index shows that the implementation of stricter emission targets, initiated particularly since the adoption of the 5th Environmental Action Programme (EAP), has effectively contributed to restraining the growth of CO$_2$ emissions across the EU. Furthermore, the index is substantially heterogeneous among the member states, distinguishing between countries that lead to emission linkage and those lagging behind. In addition, \cite{papiez2022does} evaluate the EU's energy policy effectiveness in achieving long-term decoupling and reports mixed findings. The study notes progress in detaching economic growth from emissions under the EU ETS and production-based accounting, but they find no advancement in separating economic growth from consumption-based accounting emissions in either group. 

While the above studies highlight the cyclical and long-term dynamics of emissions in response to policy, the Cohesion Policy literature has concentrated mainly on regional development objectives, which is its primary objective. Much of the literature has focused primarily on understanding its impact on regional development, specifically whether the funds have been effective in reducing regional disparities. For instance, very early on, \cite{boldrin2001inequality} pointed out that these policies tend to be more redistributive than growth-oriented, reflecting political compromises over economic efficiency. However, investments in education and human capital, which have gained a significant portion of these funds, have shown medium-term positive impacts, emphasising the critical role of human capital development in fostering long-term regional growth and convergence \citep{rodriguez2004between}. Institutional quality plays a relevant role in the effectiveness of the funds. Evidence shows that local and regional governance directly affects economic growth and the effectiveness of structural funds \citep{rodriguez2015quality, charron2014regional}. 

Although relatively few studies have explicitly focused on the effect of EU structural funds on reducing emissions, some anticipate that many of the same structural patterns that determine growth performance may also shape the trajectory of sustainable growth. In particular, some contributions emphasise the importance of regional asymmetries in the green transition. \cite{rodriguez2024green} warn that while the green transition may benefit certain regions, it can also exacerbate existing territorial inequalities, particularly affecting less-developed regions. They claim that less-developed, peri-urban, and rural regions in Southern and Eastern Europe are systematically more vulnerable, while metropolitan areas are better positioned to embrace green opportunities thanks to stronger infrastructure, skills, and institutional capacity. At the same time, the micro-foundations of the green upgrading further stress the role of \textit{regional capability-relatedness} \citep{Boschma02012017}. \citet{santoalha2021diversifying} show that regions are more likely to diversify into green technologies when related capabilities are present. They also claim that political support does not exert a strong direct effect but interacts with capabilities: regional-level support strengthens, while national-level support may moderate the role of relatedness. Consistent with this territorial heterogeneity, \citet{duenas2025regional} show that European regions experience markedly different emission dynamics under the EU ETS, with less-developed regions displaying higher volatility and more uncertain decarbonisation pathways than their richer counterparts, suggesting that the growth–emissions nexus is mediated by regional characteristics, institutions, and historical trajectories. Complementing this evidence, \citet{nichifor2025renewable} find that in Eastern Europe, renewable investments and environmental expenditures reduce emissions on average, but their effectiveness is spatially uneven and may exhibit diminishing returns beyond policy-relevant thresholds, underscoring the need for coordinated, place-based strategies.

In summary, in light of these findings, the effectiveness of the ESI funds in driving a low-carbon transition is conditional not only on structural factors such as institutional quality and regional development levels, but also on the phase of the economic cycle. Less-developed regions, characterised by higher vulnerability and weaker absorptive capacity, may struggle to convert expenditures into immediate emission reductions, potentially exhibiting weaker or delayed effects (and even short-run increases), while more developed regions are better positioned to leverage such funds into abatement. These asymmetries align with the logic of the environmental Kuznets hypothesis, suggesting that decoupling may be problematic, as regions tend to improve their environmental performance only after reaching a certain level of wealth. Therefore, seeking economic development while reducing disparities across European regions, and at the same time avoiding environmental degradation, remains a major challenge in the EU policy framework.

Accordingly, the central hypothesis of this paper is as follows:
\begin{quote}
\noindent\textbf{Hypothesis 1}: Low‑carbon economy expenditures from the European Structural and Investment Funds have heterogeneous impacts on GHG emissions across regions at different stages of development, influencing both the long‑run trend and the magnitude of the business‑cycle component of emissions.
\end{quote}

Building on recent contributions emphasising regional asymmetries of European regions (as in \citealt{santoalha2021diversifying} and \citealt{rodriguez2024green}), it can be further expected the following:

\begin{quote}
\noindent\textbf{Hypothesis 2}: The effectiveness of low-carbon expenditures is conditional on regional development levels, with weaker---and in some cases emission-increasing---impacts in less-developed regions.
\end{quote}

To test these hypotheses, rather than focusing on total emissions, we concentrate on per capita emissions and emissions intensity (relative to regional GDP) . These indicators enhance comparability across regions by accounting for differences in population and economic scale, while also isolating environmental efficiency more effectively.

\section{Data and methods}
\label{sec:data}

We aim to analyse the impact of expenditure on GHG emissions resulting from expenditure on the thematic objective related to the low-carbon economy in the 2007-2013 and 2014-2020 ESI funds programmes (a more detailed definition of the expenditure variables is presented below). We shall use relevant covariates to control economic activity, sectoral composition, and environmental policy stringency.

\subsection{Data}

\subsubsection*{Emissions data}

We focus on emissions data reported by the Emissions Database for Global Atmospheric Research (EDGAR), managed by the Joint Research Centre of the European Commission. EDGAR provides quantitative estimates of global GHG emissions by country and sector, including CO$_2$, CH$_4$, N$_2$O, and F-gases, using a methodology aligned with Intergovernmental Panel on Climate Change (IPCC) guidelines \citep{EDGAR_report}. Specifically, we use European regional data at the NUTS 2 level of aggregation, focusing on aggregated GHG emissions and their breakdown into CO$_2$, CH$_4$, and N$_2$O emissions.

\subsubsection*{ESI funds and eligibility}

We use data from the Cohesion Open Data Platform, which provides a granular view of the thematic content financed during the 2007-2013 and 2014-2020 programmes. The datasets for both periods encompass cumulative categorisation data on projects as reported in the European Regional Development Fund (ERDF), Cohesion Fund (CF), and European Social Fund (ESF) programme final implementation reports as presented in the closure documentation.\footnote{For 2007-2013 raw categorisation data see: \url{https://cohesiondata.ec.europa.eu/2007-2013-Categorisation/2007-2013-Cohesion-categorisation-raw-data-from-cl/qt3c-wiyg/about_data}.\\For 2014-2020 raw categorisation data see: \url{https://cohesiondata.ec.europa.eu/2014-2020-Categorisation/ESIF-2014-2020-ERDF-CF-ESF-raw-categorisation/xe4p-7b9q/about_data}.} These data contain the volume of EU support allocated to projects alongside a combination of categorisation dimension codes, including priority themes, the form of finance, the territorial context, location (using the Nomenclature of Territorial Units for Statistics, NUTS), and the economic dimension. We aggregate investments at NUTS level 2 (NUTS 2), as defined in version 2021. To do this, we consider all related investments, regardless of their original level of geographic specificity in the original database---be it national, NUTS-1, or even more granular at NUTS-3.

\begin{figure}[h!]
\centering
\includegraphics[width=0.98\linewidth]{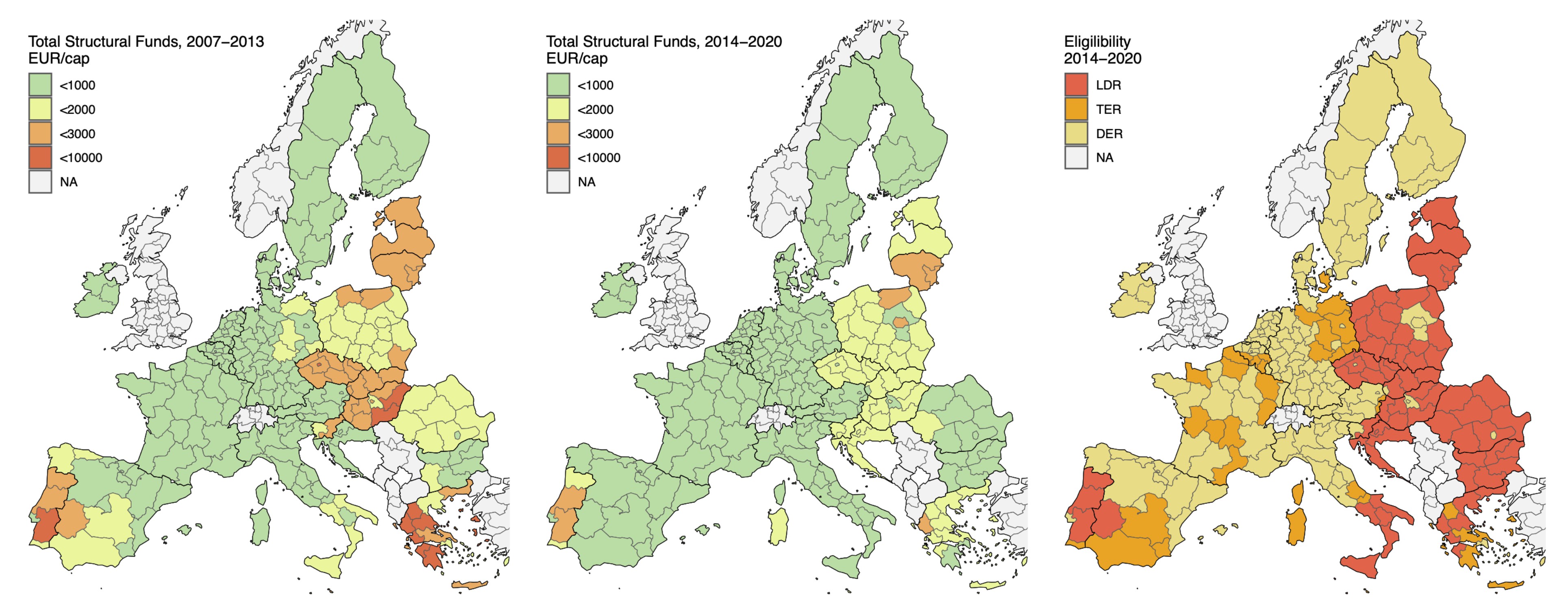}
\caption{Maps depicting the distribution of total ESI funds for the 2007-2013 and 2014-2020 programmes. Left: the total funds expenditures per capita related to the 2007-2013 program. Centre: the total funds expenditures per capita related to the 2014-2020 program. Right: the regional eligibility to access ESI funds, according to the 2014-2020 programme. Expenditure values correspond to own estimations, and all maps are at NUTS 2 level.}\label{fig:total_funds}
\end{figure}

Figure~\ref{fig:total_funds} shows the calculated distribution of total ESI funds expenditures reported by region for the periods 2007-2013 and 2014-2020, along with the eligibility of the regions to access these funds. Eligibility is based on an income classification of regions, categorising them into three groups: LDR (less developed regions), TER (transition regions), and DER (developed regions). In all econometric exercises in this paper, we use this regional income categorisation fixed.\footnote{The details of the eligibility classification, according to the ESI funds assigned in 2014-2020, can be found here: \url{https://eur-lex.europa.eu/legal-content/EN/TXT/?uri=CELEX:32014D0099}.} 

The total expenditures reflect the targeted efforts to reduce economic disparities within the EU. Certain regions consistently receive higher funds according to ongoing economic development needs or projects requiring sustained investment. The disparities in the distribution of funds, consistent with eligibility, show that less developed regions in Southern and Eastern Europe generally receive higher per capita funding than Western and Northern Europe regions. 

There are also variations in the distribution of funds between programming periods, which must be explained by changes in policy focus. Clearly, the EU's structural funds programmes significantly evolved from their initial focus on procuring the catching-up of less developed regions towards a more strategic approach. In the 2007-2013 programming period, a transition towards integrating environmental and climatic considerations began, emphasising energy efficiency, renewable energies, and sustainable resource management. However, the transition to a more strategic approach was cemented during the 2014-2020 programming period, with a clear directive to channel funds towards smart, sustainable, and inclusive growth. 

The 2014-2020 funding period introduced thematic objectives (TOs), including: Research and Innovation (TO-1), Information and Communication Technology (TO-2), Competitiveness of Small and Medium-Sized Enterprises (TO-3), Low-Carbon Economy (TO-4), Climate Change Adaptation, Risk Prevention, and Management (TO-5), Environmental Protection and Resource Efficiency (TO-6), Sustainable Transport and Removing Bottlenecks in Key Network Infrastructures (TO-7), Employment and Labour Mobility (TO-8), Social Inclusion and Combating Poverty (TO-9), Education, Training, and Vocational Training for Skills and Lifelong Learning (TO-10), and Institutional Capacity Building and Efficient Public Administration (TO-11). 

\subsubsection*{Low-carbon economy funds}

We define as ``climate action funds'' all expenditures aimed at reducing emissions in both programming periods. Despite thematic changes between the two periods, the raw categorisation data allows for the creation of a comparative set of expenditure objectives to ensure comparability and coherence in our analysis.

For the 2014-2020 programme, the low-carbon expenditures are straightforward to determine since there is a categorical variable related to thematic objective 4 (TO-4), labelled as ``low-carbon economy''. The aim of the expenditure can be tracked by the reported ``intervention field'' in the data. Specifically, related to TO-4, the intervention fields include renewable energy (wind/solar/biomass), energy efficiency renovation of public infrastructure, cycle tracks and footpaths, among others (see Appendix A1).

For the 2007-2013 programme, there is no categorical variable explicitly associated with the low-carbon economy as in the 2014-2020 programme. However, the raw categorisation data describe the ``priority'' related to the reported expenditures, serving as a proxy for the aim of the expenditure. We use this information to create a low-carbon expenditures variable for this period. Priority labels related to this period include cycle tracks, intelligent transport systems, renewable energy (wind/solar/biomass/hydroelectric), energy efficiency, co-generation, energy management, and promotion of clean urban transport. The priority codes are reported in Appendix A1.

A critical aspect of the ESI funds expenditure data is the absence of detailed annual expenditure information at the regional level. While reported expenditures can be easily tracked at the regional level, the reports correspond to the entire programming period. To our knowledge, there is no disaggregated time-varying consolidated official information available. Given this constraint, we built an annually growing cumulative investment series for each region and programme. This implies that, within each programming period, the total per capita expenditure is allocated progressively across the corresponding years. While this introduces a smoothing assumption, it allows us to preserve the annual variation in emissions and covariates. Therefore, considering the capital-intensive nature of the funded projects, we argue that their environmental effects are plausibly associated with the cumulative stock of investments. We further validate this modelling choice through robustness checks, including alternative cross-sectional estimations presented in Appendix A2.

\subsubsection*{Supplementary covariates}

Economic variables were obtained from the ARDECO (the Annual Regional Database of the European Commission's Directorate General for Regional and Urban Policy), which contains annual data at the regional level and is maintained and updated by the Joint Research Centre of the European Commission. We use variables such as the real gross domestic product, the population, and the real gross value added for all industry sectors according to NACE Rev. 2 definitions. These sectors are A: Agriculture, Forestry and Fishing; B--E: Industry; F: Construction; G--J: Wholesale, Retail, Transport, Accommodation \& Food Services, Information and Communication; K--N: Financial \& Business Services; and O--U: Non-market Services. The choice of these variables is motivated by their relevance to understanding how economic activity is related to emissions. In particular, we include industry-level GVA to control direct sources of emissions and to better capture regional emission profiles, given that the participation of industries varies significantly across European regions. 

To assess the effects of environmental policies on emissions, we use the EPS index \citep{EPS_oecd}. This index, developed by \cite{brunel2013measuring} and further refined by \cite{botta2014measuring}, provides a standardised composite measure incorporating both market and non-market-based policies, covering 27 OECD countries between 1990 and 2020. 

Finally, to determine the political ideology at the regional level, we calculate the regional weighted average of political ideologies using the results of European subnational election data. To measure political ideology, we use the ``left/right'' dimension variable reported by ParlGov \citep{ParlGov} and complemented by CHES-Europe \citep{CHES}, which takes values in the interval 0-10, with values close to zero representing extreme left ideology and values close to 10 representing extreme right ideology. We use EU-NED \citep{schraff2023european}, which reports sub-national election data across European countries over the last three decades at the party level. We assign each party at the regional level an ideology by combining information from ParlGov and CHES-Europe. These databases are integrated within the Party Facts platform, which facilitates the incorporation of party-level data.\footnote{The data are available at: \url{https://www.chesdata.eu/ches-europe} and \url{https://www.parlgov.org/}. Also see: \url{https://partyfacts.herokuapp.com/}.} We include this variable to provide additional controls for political ideology. While our focus is not on analysing how politics affects emissions, this is a growing area of interest in political science. For instance, \cite{jahn2021quick} highlights regional variations in the impact of populist governance on environmental policies, showing that right-wing populist administrations in North Western and Eastern Europe are linked to heightened GHG emissions, whereas their left-wing counterparts in Southern Europe are associated with decreases in emissions.

\subsubsection*{Distribution of emissions and low-carbon economy funds}

\begin{figure}[h!]
\centering
\includegraphics[width=0.98\linewidth]{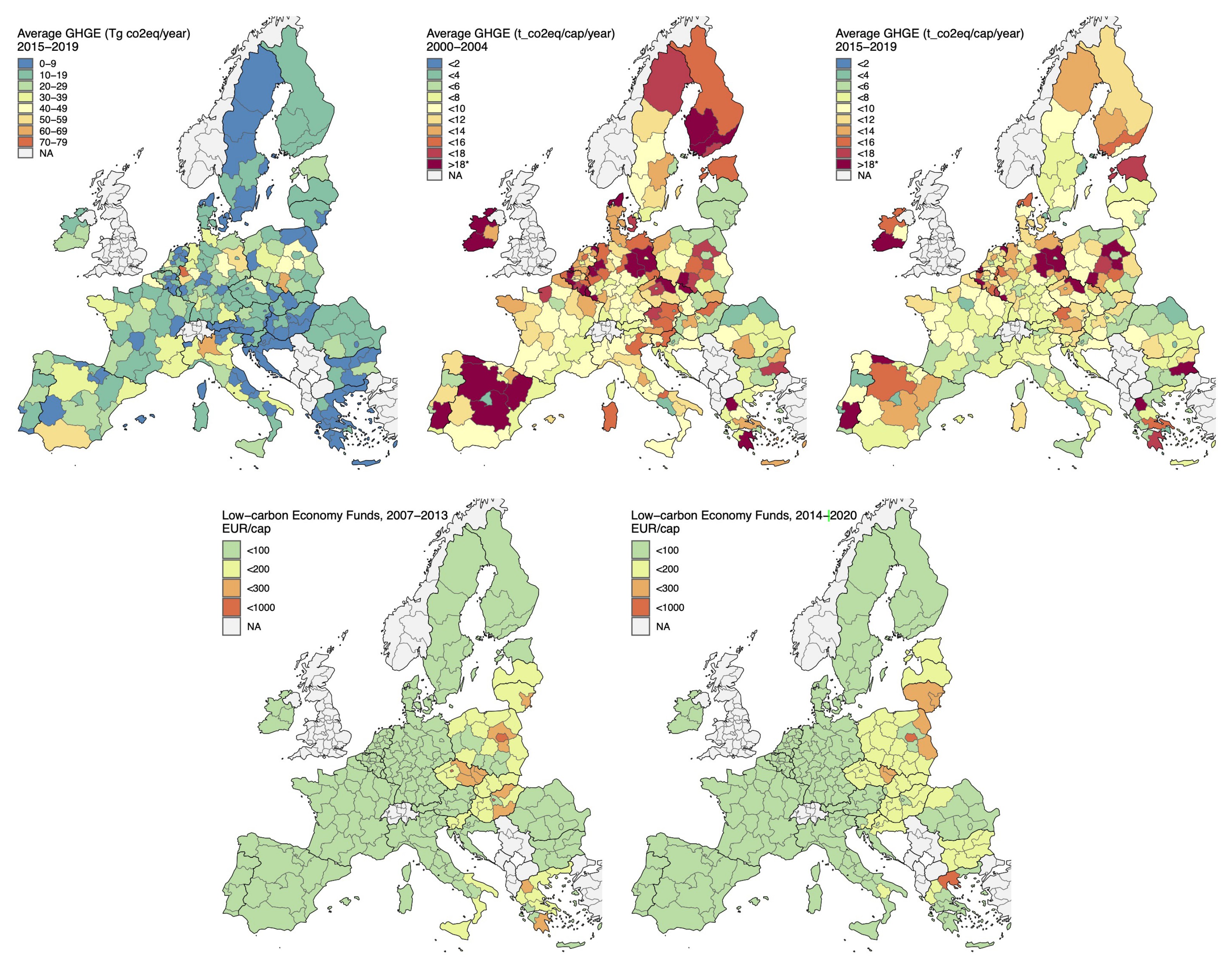}
\caption{Maps depicting GHG emission indicators, Funding Eligibility and Total Low-Carbon Economy expenditures allocation for the 2007-2013 and 2014-2020 Programmes. These maps illustrate the geographical distribution of regions according to their emissions and total expenditure within each ESI funding period. All maps are at NUTS 2 level.}\label{fig:emissions_funds}
\end{figure}

Figure~\ref{fig:emissions_funds} presents a set of maps illustrating different facets of the environmental impact and distribution of the ESI funds. In the upper part, three maps show average GHG emissions in shades of blue to red, where blue denotes lower levels and red denotes higher levels. The first map, on the left, represents the average emissions for 2015-2019 in teragrams of CO$_2$ equivalent per year (Tg CO$_2$eq/year). The other two maps compare the average annual per capita emissions for 2000-2004 and 2015-2019, respectively. These maps show that, on average, towards the end of the 2010s, most European regions emitted less than 50 teragrams of CO$_2$. However, there is significant heterogeneity in per capita emissions, highlighting areas of intense economic activity or low population density.\footnote{Sparsely populated regions may have high per capita emissions despite relatively moderate total emissions. This is due to emission-intensive activities, like agriculture and energy production, concentrated in some low-density areas.} Comparing the early 2000s and late 2010s, it is evident that many regions with initially high emissions tended to reduce them (for instance, the Castile region in Spain), while some regions maintained or increased their per capita emission levels (for instance, the Masovian region in Poland).

The lower maps in Figure~\ref{fig:emissions_funds} show the calculated funds allocated to promote a low-carbon economy during 2007-2013 and 2014-2020 programmes, with the amounts received per capita in colour scales ranging from less than 100 euros to almost 1,000 euros per person. In both periods, most of the funds went to LDRs, with significant differences in the allocated amount, in agreement with the results presented in Figure~\ref{fig:total_funds}. 

As a preliminary analysis, we estimate the nexus between expenditure on low-carbon economy funds and total GHG emissions using the following regional-level model:
\begin{equation}
\label{eq:model0} 
Y_{i,t} = \beta + \alpha_1 \: s_{i,t-\tau}^{07-13} + \alpha_2 \: s_{i,t-\tau}^{14-20} +  
\gamma_i + \gamma_t + \varepsilon_{i,t} \;;
\end{equation}
where $Y_{i,t}$ is an emission indicator for region $i$ at time $t$, and $s_{i,t-\tau}^{07-13}$ and $s_{i,t-\tau}^{14-20}$ are the cumulative yearly programme expenditures per capita for the periods 2007-2013 and 2014-2020, respectively, with $\tau$ denoting lags to acknowledge potential delayed effects. We include fixed effects for location ($\gamma_i$) and time ($\gamma_t$) to control for time-invariant unobserved heterogeneity across regions as well as uniform temporal shocks, such as those associated with environmental directives instituted at the European Union level. Finally, the term $\varepsilon_{i,t}$ is the estimation residual. 

In the left-hand side of Eq.~\eqref{eq:model0}, the term $Y_{i,t}$ serves as a reference notation for different emission variables. Specifically, we use four different dependent variables based on the total emissions $E_{i,t}$ of region $i$ at time $t$. These are the levels of emissions per capita (in log), defined as GHGEpc$_{i,t}$=$E_{i,t}/pop_{i,t}$; the emission intensity relative to the GDP, defined as GHGEi$_{i,t}$=$E_{i,t}/GDP_{i,t}$; and the growth rates of GHGEpc$_{i,t}$ and GHGEi$_{i,t}$. 

The estimation results are presented in Table~\ref{tb:model0}. We find that expenditures have a statistically significant positive relationship with per capita emissions, as seen in models 1 and 2, indicating a perverse effect against the intended impact of such investments. When examining the growth rates of emissions per capita, a negative relationship appears with the three-year lag in spending (model 3). However, in model 4, which accounts for a five-year lag, this relationship is not statistically significant. Regarding emission intensity, we find that low-carbon expenditures are linked to decreases in this measure, as demonstrated in models 5 and 6. At the same time, the growth rate of emission intensity does not show a significant effect related to the expenditures. 

\begin{table}[h!]
\begin{center}
\scriptsize 
\caption{Panel fixed-effects estimation results ($year>=2007$)}\label{tb:model0}\vspace{3pt}
\resizebox{0.8\textwidth}{!}{\renewcommand{\arraystretch}{1.1}
\begin{tabular}{l cc cc cc cc} 
\toprule 
 & \multicolumn{4}{c}{GHGEpc} & \multicolumn{4}{c}{GHGEi} \\ 
\cmidrule(lr){2-5}
\cmidrule(lr){6-9}
 & \multicolumn{2}{c}{Levels (log)} & \multicolumn{2}{c}{Growth rates}  & \multicolumn{2}{c}{Levels (log)} & \multicolumn{2}{c}{Growth rates} \\ 
\cmidrule(lr){2-3}
\cmidrule(lr){4-5}
\cmidrule(lr){6-7}
\cmidrule(lr){8-9}
 & (1) & (2) & (3) & (4) & (5) & (6) & (7) & (8)\\ 
\midrule
 Low Carbon Exp. pc$^{07-13}_{t-3}$ & 0.004 &  & $-$0.081$^{**}$ &  & $-$0.130$^{**}$ &  & $-$0.042 &  \\ 
  & (0.063) &  & (0.036) &  & (0.064) &  & (0.038) &  \\ 
  & & & & & & & & \\ 
 Low Carbon Exp. pc$^{14-20}_{t-3}$ & 1.169$^{***}$ &  & $-$0.008 &  & $-$0.438$^{***}$ &  & $-$0.009 &  \\ 
  & (0.091) &  & (0.053) &  & (0.093) &  & (0.055) &  \\ 
  & & & & & & & & \\ 
 Low Carbon Exp. pc$^{07-13}_{t-5}$ &  & 0.149$^{**}$ &  & 0.052 &  & $-$0.123$^{**}$ &  & 0.028 \\ 
  &  & (0.062) &  & (0.035) &  & (0.063) &  & (0.037) \\ 
  & & & & & & & & \\ 
 Low Carbon Exp. pc$^{14-20}_{t-5}$ &  & 1.633$^{***}$ &  & 0.067 &  & $-$0.601$^{***}$ &  & 0.091 \\ 
  &  & (0.140) &  & (0.080) &  & (0.142) &  & (0.084) \\ 
\hline
Observations & 3,872 & 3,872 & 3,872 & 3,872 & 3,872 & 3,872 & 3,872 & 3,872 \\ 
\bottomrule
\multicolumn{9}{p{16cm}}{\textit{Notes:} NUTS 2 cluster robust standard errors are in parentheses. Significance level: $^{*}$p$<$0.1; $^{**}$p$<$0.05; $^{***}$p$<$0.01}\\
\end{tabular}}
\end{center}
\end{table}

These mixed results suggest that the econometric specification might be masking underlying factors that could restrict the effectiveness of low-carbon funds. One relevant limiting factor is the persistent disparities among European regions. Indeed, the development level of regions is a key factor influencing the effectiveness of the ESI funds \citep{boldrin2001inequality, rodriguez2004between}. However, many aspects related to the heterogeneity of European regions can explain the mixed results. For instance, less developed regions might use these funds to build sustainable infrastructure to reduce emissions, while more developed regions, having already made progress in emissions efficiency, might experience a more moderate impact from these funds. Therefore, the empirical strategy must consider that short-term outcomes may differ from longer-term impacts. This consideration is crucial for environmental policy because it suggests that sustainability investments may be misallocated, leading to unintended or unfavourable effects on emissions.

\subsection{Methodology}

The research on decoupling emissions from economic activity primarily hinges on analysing the emissions-GDP elasticity, which measures how emissions change with GDP and helps understand the link between economic growth and environmental impact. The central model is as follows: 
\begin{equation} 
\label{eq:elast}
\Delta e_{t} = \alpha + \beta \: \Delta y_{t} + u_{t} \;;
\end{equation}
where $e_{t}$ represents the log of total emissions at time $t$, $y_{t}$ denotes the log of real GDP, $\Delta$ denotes these first differences operator, and $u_{t}$ represents the error term. The parameters $\alpha$ and $\beta$ capture the intercept and the elasticity of emissions with respect to GDP, respectively. 

However, as \cite{cohen2018long} pointed out, this approach has some important caveats. Looking at the growth of emissions in relation to real GDP growth for the world's 20 largest emitters, the authors showed a positive emissions-output elasticity across all countries, suggesting a strong correlation between economic activity and emissions. However, by decomposing emissions and real GDP into their trend and cyclical components, they show clearer evidence of decoupling in richer nations, particularly in European countries. The trend components of this analysis, which focus on long-term movements rather than short-term fluctuations, indicate that while cyclical components are still closely tied to GDP growth, the trend components are not. For several countries, including Italy, the United Kingdom, and Germany, the trend elasticities for emissions are either close to zero or negative, suggesting that the long-term component of emissions growth has decoupled from GDP growth \citep{cohen2018long}.

Our methodology shifts from this type of analysis, which primarily assesses the impact of GDP growth on emissions, to a comparative panel data analysis focusing on the effects of expenditure in the low-carbon economy. We specifically highlight the role of the ESI funds in fostering low-carbon economies to curb emissions. Therefore, it is central in our methodology to decompose time series variables, such as emissions, GDP, and GVA, into their cyclical and trend components for each region independently. Then, we consider these time series in two different panel data model frameworks: regional trends and regional cycles of emissions. In this way, we analyse the difference between long-term trends and cyclical effects of the impact of low-carbon economy funds on GHG emissions.

In the first phase of our methodology, we apply the Hodrick-Prescott (HP) filter to each region's emissions per capita and emissions intensity time series to separate them into their corresponding cyclical and trend components. This method allows us to distinguish short-term fluctuations associated with the economic cycle from the long-term trajectory that reflects structural changes and persistent trends in emissions. Therefore, the total emissions time series of the region $i$, $e_{i,t}$, is decomposed in its trend component $e_{i,t}^{\{trend\}}$ and its cycle component $e_{i,t}^{\{cycle\}}$. 

After decomposing the GHGEpc and GHGEi time series, we independently analyse the panel data of regional emission trends and cycles, taking into account differences in regional development levels. The model for the trend and cycle components of emissions is defined as follows:
\begin{equation}\label{eq:model}
\begin{split}
e_{i,t}^{\{trend,cycle\}} = & \:\alpha + \beta \: y_{i,t}^{\{trend,cycle\}} + 
\theta_1 \: s_{i,t-\tau}^{07-13} + 
\sum_{l_i}\phi_{l_i} \: DL_i \times s_{i,t-\tau}^{07-13} + \\
& \theta_2 \: s_{i,t-\tau}^{14-20} + 
\sum_{l_i}\psi_{l_i} \: DL_i \times s_{i,t-\tau}^{14-20} + 
Z_{i,t-\tau} + \gamma_i + \gamma_t + \varepsilon_{i,t} \;;
\end{split}
\end{equation}
where the superscript $\{trend,cycle\}$ indicates that we estimate the model separately for the trend and cycle components of the series. Specifically, $e_{i,t}^{\{\cdot\}}$ denotes either the trend or the cycle component of the logarithm of an emissions indicator (GHGEpc or GHGEi)\footnote{Using these relative measures offers several advantages: first, per‑capita emissions normalise for population size, allowing us to compare each region's carbon footprint regardless of its scale; second, emissions intensity controls for economic scale, highlighting how efficiently regions produce output per unit of carbon. This enhances comparability across diverse regions and reduces heteroskedasticity in panel regressions. More importantly, it allows us to better isolate the impact of low‑carbon investments on environmental efficiency and avoid confounding with demographic or economic differences.} for region $i$ at time $t$, and $y_{i,t}^{\{\cdot\}}$ is the corresponding component of the logarithm of GDP. The terms $s_{i,t-\tau}^{07-13}$ and $s_{i,t-\tau}^{14-20}$ refer to the cumulative yearly expenditures of the 2007–2013 and 2014–2020 programmes, respectively, lagged by $\tau$ years. Regional development levels are introduced via the dummy variable $DL_i$, with $l_i = \{\text{LDR}, \text{TER}, \text{DER}\}$ representing less developed, transition, and developed regions, respectively.

The vector $Z_{i,t-\tau}$ includes additional controls, such as a dummy for environmental policy stringency ($EPS$)\footnote{Following \cite{naqvi2021decoupling}, we use a dummy equal to one when the EPS index is greater than or equal to 2.5.} and an indicator of average political ideology ($LR$), based on local election outcomes. The model also incorporates region fixed effects ($\gamma_i$) and year fixed effects ($\gamma_t$) to control for time-invariant regional characteristics and common shocks across all regions. As a robustness check, we replace GDP with the set of GVA variables for all economic sectors to better account for the role of regional productive structures in explaining GHG emissions.

While our specification includes key relevant variables, we acknowledge that endogeneity may still pose concerns. For instance, one might argue that environmental regulations or improvements in energy efficiency could, in principle, influence regional output and investment decisions. Moreover, the allocation of ESI funds is determined based on regional eligibility criteria related to development levels, making the investment variable exogenous to current emission levels. By interacting expenditures with regional development levels, we account for heterogeneous effects while controlling for potential confounding factors. Nonetheless, one could still argue that regions with higher emissions may have received more funds, representing a case of reverse causality or omitted variable bias. 

For these reasons, we have included a five-year lag for all explanatory variables, ensuring that the predictors precede the observed emissions outcomes by a meaningful temporal distance. Therefore, our specification mitigates---though does not fully eliminate---simultaneity concerns and relies on the timing assumption that lagged expenditures impact emissions. A plausible strategy would have been to implement instrumental variables; however, given the nature of the impact of low-carbon investments---which take time to materialise---it is difficult to identify strong instruments.\footnote{As a robustness check, in Appendix A3, we implemented 2SLS estimations using structural and institutional variables (e.g. labour productivity, hours worked per capita, consumption of fixed capital, compensation of employees) as potential instruments. Despite these instruments proved to be valid, they are not strong enough to support a reliable specification, specifically the Kleibergen-Paap rk Wald F statistic is well below conventional thresholds. Thus, these IV results must be interpreted with caution.}

\section{Results}
\label{sec:results}

\subsection{Effect on emissions trend}

Table~\ref{tb:model_trend} presents the estimation results of Eq.~\eqref{eq:model} using as the dependent variable the trends for GHGE per capita or GHGE intensity. The key independent variables are low-carbon expenditures per capita from the 2007-2013 and 2014-2020 programmes, each introduced with a five-year lag. The heterogeneous impacts of regional development levels are captured by interacting these investment variables with dummy variables for transition (TER) and less-developed regions (LDR), using developed regions (DER) as the reference category. 
\begin{table}[h!]
\begin{center}
\scriptsize 
\caption{Fixed effect estimation results on GHG emissions trends ($year>=2007$)}\label{tb:model_trend}\vspace{3pt}
\resizebox{0.7\textwidth}{!}{\renewcommand{\arraystretch}{1.1}
\begin{tabular}{lcccccc} 
\toprule 
 & \multicolumn{6}{c}{\textit{Dependent variable:}} \\ 
\cmidrule{2-7} 
 & \multicolumn{3}{c}{GHGE per capita} & \multicolumn{3}{c}{GHGE intensity} \\ 
\cmidrule(lr){2-4}
\cmidrule(lr){5-7}
& (1t) & (2t) & (3t) & (4t) & (5t) & (6t)\\ 
\midrule
Low Carbon Exp. pc$^{07-13}_{t-5}$  & 0.039 & -0.024 & 0.065 & -0.383*** & -0.073 & -0.037 \\
   & (0.177) & (0.143) & (0.102) & (0.122) & (0.154) & (0.126) \\
Low Carbon Exp. pc$^{14-20}_{t-5}$  & -0.182 & -0.446 & 0.020 & -1.117 & -0.763 & -0.458 \\
   & (0.586) & (0.733) & (0.680) & (0.848) & (0.673) & (0.641) \\
Low Carbon Exp. pc$^{07-13}_{t-5}$$\times$TER  & -1.081* & -0.491 & -0.638 & 0.065 & -0.337 & -0.375 \\
   & (0.644) & (0.790) & (0.862) & (0.604) & (0.517) & (0.567) \\
Low Carbon Exp. pc$^{14-20}_{t-5}$$\times$TER  & -1.200 & -0.409 & -1.371 & 0.426 & -0.608 & -1.005 \\
   & (1.613) & (1.585) & (1.561) & (1.461) & (1.510) & (1.400) \\
Low Carbon Exp. pc$^{07-13}_{t-5}$$\times$LDR  & 0.611*** & 0.548*** & 0.439** & 0.458** & 0.383 & 0.327 \\
   & (0.226) & (0.200) & (0.185) & (0.219) & (0.259) & (0.248) \\
Low Carbon Exp. pc$^{14-20}_{t-5}$$\times$LDR  & 1.315** & 1.160* & 0.728 & 0.668 & 0.784 & 0.497 \\
   & (0.587) & (0.681) & (0.644) & (0.810) & (0.632) & (0.611) \\
GDP$_{t-5}$  &  & 0.238*** &  &  & -0.462*** &  \\
   &  & (0.065) &  &  & (0.102) &  \\
EPS$_{t-5}$  &  & 0.000 & 0.001 &  & 0.001 & 0.001 \\
   &  & (0.007) & (0.006) &  & (0.007) & (0.007) \\
Political Ideology$_{t-5}$  &  & 0.014 & 0.007 &  & -0.011 & -0.009 \\
   &  & (0.010) & (0.009) &  & (0.016) & (0.014) \\
GVA$_{A,t-5}$  &  &  & 0.065* &  &  & 0.034 \\
   &  &  & (0.034) &  &  & (0.038) \\
GVA$_{B-E,t-5}$  &  &  & 0.242*** &  &  & -0.042 \\
   &  &  & (0.047) &  &  & (0.075) \\
GVA$_{F,t-5}$  &  &  & -0.135*** &  &  & -0.153*** \\
   &  &  & (0.041) &  &  & (0.057) \\
GVA$_{G-J,t-5}$  &  &  & 0.129 &  &  & -0.200 \\
   &  &  & (0.091) &  &  & (0.132) \\
GVA$_{K-N,t-5}$  &  &  & -0.362*** &  &  & -0.391*** \\
   &  &  & (0.084) &  &  & (0.100) \\
GVA$_{O-U,t-5}$  &  &  & 0.170 &  &  & 0.310** \\
 &  &  & (0.108) &  &  & (0.134) \\
\hline
Observations & 3,872 & 3,408 & 3,408 & 3,872 & 3,408 & 3,408 \\
\bottomrule
\multicolumn{7}{p{13cm}}{\textit{Notes:} Robust standard errors clustered at the NUTS 2 level are reported in parentheses. Significance level: $^{*}$p$<$0.1; $^{**}$p$<$0.05; $^{***}$p$<$0.01.}\\
\end{tabular}}
\end{center}
\end{table}

The estimates for per capita emissions (columns 1t–3t) show that in developed regions, the reference group, the coefficients on low-carbon expenditures are small and not statistically significant across all models. In contrast, the interaction terms for less developed regions are consistently positive and significant for the 2007–2013 programme, suggesting that greater expenditure is associated with increases in long-run emissions per capita. The interaction terms for the 2014–2020 period are also positive, although they are significant only in models 1t and 2t. For transition regions, the interaction coefficients are negative in most cases, but not statistically significant, except marginally in model 1t for 2007–2013 expenditures.

Regarding emissions intensity (columns 4t-6t), we observe a statistically significant negative effect of 2007–2013 expenditures in developed regions only in model 4t, suggesting improved emissions efficiency. However, for less developed regions, the corresponding interaction terms are again positive and significant in model 4t. As in the GHGEpc models, results for transition regions remain statistically not significant across all models.

These results support H1, as the effects of low-carbon expenditures are heterogeneous across development levels and emission indicators. Moreover, the consistently positive and significant interactions for LDR alongside weak or null effects for DER/TER provide evidence in favour of H2.

Control variables, particularly GDP and sectoral GVA, exhibit expected significant relationships with emissions trends. Higher GDP is linked positively with emissions per capita but negatively with emissions intensity, highlighting the scale versus efficiency effects inherent in emissions analysis. Sectoral GVA variables reveal heterogeneity across economic activities. For instance, GVA in sectors B–E (industry) is positively correlated with per capita emissions, while GVA in services (sectors K–N) shows a significant negative association with both emissions indicators. These results underline the importance of regional economic structure in shaping emissions trajectories.

As regards institutional and political controls, we find that environmental policy stringency and political ideology show no statistically significant effects. The small positive coefficients on EPS suggest that national-level measures may not strongly influence regional emissions trends in per capita or intensity terms. Similarly, political orientation, as proxied by the left-right index, does not exhibit consistent effects.

\subsection{Effect in the emissions cycle}

Table~\ref{tb:model_cycle} presents the estimation results of Eq.~\eqref{eq:model}, using the cyclical components of GHG emissions per capita and emissions intensity as dependent variables.
\begin{table}[h!]
\begin{center}
\scriptsize 
\caption{Fixed effect estimation results on GHG emissions cycles ($year>=2007$)}\label{tb:model_cycle}\vspace{3pt}
\resizebox{0.8\textwidth}{!}{\renewcommand{\arraystretch}{1.1}
\begin{tabular}{lcccccc} 
\toprule 
 & \multicolumn{6}{c}{\textit{Dependent variable:}} \\ 
\cmidrule{2-7} 
 & \multicolumn{3}{c}{GHGE per capita} & \multicolumn{3}{c}{GHGE intensity} \\ 
\cmidrule(lr){2-4}
\cmidrule(lr){5-7}
& (1c) & (2c) & (3c) & (4c) & (5c) & (6c)\\ 
\midrule
Low Carbon Exp. pc$^{07-13}_{t-5}$  & -0.017 & -0.055** & -0.065** & 0.027 & 0.004 & -0.002 \\
   & (0.021) & (0.028) & (0.027) & (0.039) & (0.030) & (0.030) \\
Low Carbon Exp. pc$^{14-20}_{t-5}$  & -0.128 & -0.189 & -0.200 & -0.092 & -0.168 & -0.180 \\
   & (0.230) & (0.235) & (0.206) & (0.249) & (0.234) & (0.224) \\
Low Carbon Exp. pc$^{07-13}_{t-5}$$\times$TER  & -0.166 & -0.130 & -0.131 & -0.107 & 0.048 & 0.043 \\
   & (0.139) & (0.117) & (0.114) & (0.173) & (0.088) & (0.093) \\
Low Carbon Exp. pc$^{14-20}_{t-5}$$\times$TER  & 0.079 & -0.096 & -0.077 & -0.358 & -0.664 & -0.662 \\
   & (0.746) & (0.738) & (0.704) & (0.817) & (0.720) & (0.725) \\
Low Carbon Exp. pc$^{07-13}_{t-5}$$\times$LDR  & -0.100** & -0.110** & -0.130*** & -0.078 & -0.096** & -0.104** \\
   & (0.050) & (0.048) & (0.047) & (0.054) & (0.042) & (0.041) \\
Low Carbon Exp. pc$^{14-20}_{t-5}$$\times$LDR  & 0.198 & 0.292 & 0.333 & 0.084 & 0.221 & 0.239 \\
   & (0.238) & (0.240) & (0.217) & (0.253) & (0.234) & (0.224) \\
GDP$_{t-5}$  &  & -0.251*** &  &  & 0.082** &  \\
   &  & (0.037) &  &  & (0.035) &  \\
EPS$_{t-5}$  &  & 0.010*** & 0.011*** &  & 0.012*** & 0.012*** \\
   &  & (0.003) & (0.003) &  & (0.003) & (0.003) \\
Political Ideology$_{t-5}$  &  & 0.005 & 0.003 &  & 0.003 & 0.003 \\
   &  & (0.003) & (0.003) &  & (0.003) & (0.003) \\
GVA$_{A,t-5}$  &  &  & -0.012 &  &  & -0.012 \\
   &  &  & (0.009) &  &  & (0.007) \\
GVA$_{B-E,t-5}$  &  &  & -0.011 &  &  & -0.003 \\
   &  &  & (0.018) &  &  & (0.018) \\
GVA$_{F,t-5}$  &  &  & -0.044*** &  &  & 0.017 \\
   &  &  & (0.014) &  &  & (0.017) \\
GVA$_{G-J,t-5}$  &  &  & 0.030 &  &  & 0.080** \\
   &  &  & (0.033) &  &  & (0.032) \\
GVA$_{K-N,t-5}$  &  &  & -0.044 &  &  & 0.024 \\
   &  &  & (0.041) &  &  & (0.038) \\
GVA$_{O-U,t-5}$  &  &  & -0.203*** &  &  & -0.058 \\
 &  &  & (0.041) &  &  & (0.042) \\
\hline
Observations & 3,872 & 3,408 & 3,408 & 3,872 & 3,408 & 3,408 \\
\bottomrule
\multicolumn{7}{p{13cm}}{\textit{Notes:} Robust standard errors clustered at the NUTS 2 level are reported in parentheses. Significance level: $^{*}$p$<$0.1; $^{**}$p$<$0.05; $^{***}$p$<$0.01}\\
\end{tabular}}
\end{center}
\end{table}

The estimates for per capita emissions cycles (columns 1c–3c) show that in developed regions, the direct effect of low-carbon expenditures is negative and statistically significant for the 2007–2013 programme in models 2c and 3c, suggesting that these investments helped smooth short-term fluctuations. Notably, the interaction terms for less-developed regions are also negative and statistically significant across all three models for the same programming period. This indicates that, despite the positive long-term trend effect observed in less-developed regions (see Table~\ref{tb:model_trend}), low-carbon investments during the 2007–2013 programme contributed to dampening short-term emissions volatility. The coefficients for transition regions are generally negative but not statistically significant.

As regards emissions intensity cycles (columns 4c–6c), there is no significant direct effect of low-carbon spending for developed or transition regions. However, for less-developed regions, interaction terms for the 2007–2013 programme are again negative and statistically significant (in models 5c and 6c). As in the per capita models, the effects of the 2014–2020 programme are not statistically significant for any regional group.

These findings reinforce H1 and H2, showing that low-carbon investments affect not only long-run trajectories but also the business-cycle component of emissions, with short-run impacts that vary heterogeneously across regions depending on their level of development.

Lagged GDP is negatively associated with per capita emissions cycles and positively with emissions intensity cycles, consistent with short-run adjustments in the emissions-output relationship. Sectoral GVA variables display heterogeneous effects: GVA in construction (sector F) is associated with a reduction in emissions per capita, while services (G–J) appear positively associated with emissions intensity in model 6c. Other sectors are not significantly correlated with the cyclical components of emissions.

Political ideology remains statistically not significant across specifications. However, environmental policy stringency shows a small but statistically significant positive effect in all models where it is included. This suggests that the cyclical component of emissions may co-move with the timing or enforcement of regulatory frameworks. These results are in line with recent findings for European regions, which report that emissions growth rates are highly volatile and sensitive to policy phases, particularly within the EU ETS framework \citep{duenas2025regional}.

\subsection{Effect by type of gases}

We break down emissions by specific gases, considering CO$_2$, CH$_4$, and N$_2$O, to assess the differentiated impact of low-carbon investments. We adopt the same methodological approach used for total GHG emissions: decomposing the per capita and intensity time series for each gas into trend and cycle components, and estimating the effects of expenditures on these indicators following the specification in Eq.\eqref{eq:model}. Results are reported in Tables~\ref{tb:model_trend_types} and \ref{tb:model_cycle_types}, corresponding to the trend and cycle components, respectively. For readability, the estimated coefficients of control variables are omitted from the tables.

Regarding the trend of CO$_2$ emissions (upper panel of Table~\ref{tb:model_trend_types}), we find no statistically significant effects of the 2007–2013 programme for the reference group (developed regions) in terms of per capita emissions. However, the interaction terms with less developed regions are consistently positive and significant across specifications. For CO$_2$ intensity, a significant negative effect is observed in column 4 for the reference group, indicating possible efficiency gains, yet these are counteracted by a positive and significant interaction with LDR in the same column. For the 2014–2020 programme, effects on CO$_2$ per capita and intensity are generally not significant. Nonetheless, the interaction with LDR remains positive and significant in columns 1 and 2, reinforcing the earlier programming effect pattern.
\begin{table}[h!]
\begin{center}
\scriptsize 
\caption{Summary of fixed-effects estimation results on emission trends by gas type}\label{tb:model_trend_types}\vspace{3pt}
\resizebox{0.7\textwidth}{!}{\renewcommand{\arraystretch}{1.1}
\begin{tabular}{lcccccc} 
\toprule 
 & \multicolumn{6}{c}{\textit{Dependent variables by gas type}} \\ 
\cmidrule{2-7} 
 & \multicolumn{3}{c}{CO$_2$ per capita} & \multicolumn{3}{c}{CO$_2$ intensity} \\ 
\cmidrule(lr){2-4}
\cmidrule(lr){5-7}
Model & (1) & (2) & (3) & (4) & (5) & (6)\\ 
\midrule
Low Carbon Exp. pc$^{07-13}_{t-5}$  & 0.066 & -0.061 & 0.077 & -0.356*** & -0.110 & -0.025 \\
   & (0.199) & (0.154) & (0.102) & (0.125) & (0.159) & (0.123) \\
Low Carbon Exp. pc$^{14-20}_{t-5}$  & -0.300 & -0.424 & 0.172 & -1.235 & -0.740 & -0.307 \\
   & (0.523) & (0.708) & (0.707) & (0.802) & (0.643) & (0.658) \\
Low Carbon Exp. pc$^{07-13}_{t-5}$$\times$TER  & -1.164 & -0.371 & -0.561 & -0.018 & -0.217 & -0.298 \\
   & (0.715) & (0.868) & (0.958) & (0.691) & (0.591) & (0.660) \\
Low Carbon Exp. pc$^{14-20}_{t-5}$$\times$TER  & -0.726 & 0.094 & -1.216 & 0.900 & -0.105 & -0.850 \\
   & (1.728) & (1.663) & (1.675) & (1.559) & (1.592) & (1.531) \\
Low Carbon Exp. pc$^{07-13}_{t-5}$$\times$LDR  & 0.621** & 0.580** & 0.420** & 0.468* & 0.415 & 0.308 \\
   & (0.258) & (0.224) & (0.202) & (0.238) & (0.282) & (0.268) \\
Low Carbon Exp. pc$^{14-20}_{t-5}$$\times$LDR  & 1.430*** & 1.189* & 0.643 & 0.783 & 0.813 & 0.411 \\
   & (0.538) & (0.667) & (0.675) & (0.778) & (0.616) & (0.633) \\
\midrule
 & \multicolumn{3}{c}{CH$_4$ per capita} & \multicolumn{3}{c}{CH$_4$ intensity} \\ 
\cmidrule(lr){2-4}
\cmidrule(lr){5-7}
Low Carbon Exp. pc$^{07-13}_{t-5}$  & -0.115 & -0.005 & -0.045 & -0.537*** & -0.054 & -0.147 \\
   & (0.181) & (0.182) & (0.175) & (0.168) & (0.215) & (0.210) \\
Low Carbon Exp. pc$^{14-20}_{t-5}$  & 0.826 & 0.042 & -0.095 & -0.109 & -0.275 & -0.573 \\
   & (0.909) & (0.801) & (0.685) & (1.084) & (0.771) & (0.699) \\
Low Carbon Exp. pc$^{07-13}_{t-5}$$\times$TER  & -0.718** & -0.507 & -0.588 & 0.429 & -0.354 & -0.324 \\
   & (0.333) & (0.426) & (0.468) & (0.367) & (0.280) & (0.290) \\
Low Carbon Exp. pc$^{14-20}_{t-5}$$\times$TER  & -2.247** & -1.191 & -1.261 & -0.621 & -1.390 & -0.894 \\
   & (1.136) & (1.103) & (1.103) & (1.215) & (1.108) & (1.087) \\
Low Carbon Exp. pc$^{07-13}_{t-5}$$\times$LDR  & 0.515** & 0.504** & 0.523** & 0.364 & 0.339 & 0.410 \\
   & (0.201) & (0.199) & (0.211) & (0.238) & (0.271) & (0.275) \\
Low Carbon Exp. pc$^{14-20}_{t-5}$$\times$LDR  & -0.435 & 0.148 & 0.119 & -1.083 & -0.227 & -0.112 \\
   & (0.887) & (0.777) & (0.679) & (1.040) & (0.764) & (0.700) \\
\midrule
 & \multicolumn{3}{c}{N$_2$O per capita} & \multicolumn{3}{c}{N$_2$O intensity} \\ 
\cmidrule(lr){2-4}
\cmidrule(lr){5-7}
Low Carbon Exp. pc$^{07-13}_{t-5}$  & -0.115 & -0.059 & -0.058 & -0.543*** & -0.092 & -0.163 \\
   & (0.100) & (0.091) & (0.070) & (0.133) & (0.131) & (0.108) \\
Low Carbon Exp. pc$^{14-20}_{t-5}$  & 0.180 & 0.142 & 0.288 & -0.774 & -0.118 & -0.196 \\
   & (0.340) & (0.382) & (0.339) & (0.631) & (0.312) & (0.299) \\
Low Carbon Exp. pc$^{07-13}_{t-5}$$\times$TER  & -0.574*** & -0.187 & -0.140 & 0.570 & -0.088 & 0.134 \\
   & (0.206) & (0.166) & (0.160) & (0.583) & (0.364) & (0.361) \\
Low Carbon Exp. pc$^{14-20}_{t-5}$$\times$TER  & -0.969*** & -0.764* & -0.829 & 0.647 & -1.049** & -0.442 \\
   & (0.369) & (0.440) & (0.527) & (0.756) & (0.411) & (0.659) \\
Low Carbon Exp. pc$^{07-13}_{t-5}$$\times$LDR  & 0.536*** & 0.347** & 0.374** & 0.380** & 0.172 & 0.267 \\
   & (0.178) & (0.144) & (0.150) & (0.193) & (0.191) & (0.176) \\
Low Carbon Exp. pc$^{14-20}_{t-5}$$\times$LDR  & 1.358*** & 0.491 & 0.315 & 0.726 & 0.089 & 0.089 \\
   & (0.424) & (0.377) & (0.355) & (0.619) & (0.309) & (0.305) \\
\bottomrule
\multicolumn{7}{p{13cm}}{\textit{Notes:} Estimated coefficients of control variables are omitted. Model specifications correspond to those in Table~\ref{tb:model_trend}. Robust standard errors clustered at the NUTS 2 level are reported in parentheses. Significance levels: $^{*}$p$<$0.1; $^{**}$p$<$0.05; $^{***}$p$<$0.01.}\\
\end{tabular}}
\end{center}
\end{table}

Turning to CH$_4$ emissions (middle panel of Table~\ref{tb:model_trend_types}), the 2007–2013 programme does not significantly affect per capita emissions in the control group. There is a negative and significant interaction with transition regions in column 1. In contrast, interactions with LDR are consistently positive and significant. For CH$_4$ intensity, a significant negative impact is observed for the reference group in column 4, but not for the interactions with TER or LDR. Under the 2014–2020 programme, we find no statistically significant effects on either CH$_4$ per capita or intensity, across all regional groups.

Finally, in the case of N$_2$O emissions (lower panel of Table~\ref{tb:model_trend_types}), the 2007–2013 programme shows a significant reduction in per capita emissions for TER in column 1, while interaction terms for LDR are again positive and significant throughout. Regarding N$_2$O intensity, a negative and significant coefficient is observed for the reference group in column 4, but positive and significant interactions with LDR offset this. For the 2014–2020 programme, the interaction with TER is negative and statistically significant in both column 1 and 2 (N$_2$O per capita) and column 5 (N$_2$O intensity). The interaction with LDR remains positive and significant only in column 1, in line with the pattern of increased emissions observed in the earlier programme.

As regards the cycle component of CO$_2$ emissions (upper panel of Table~\ref{tb:model_cycle_types}), the 2007–2013 programme is associated with a statistically significant reduction in per capita emissions for the reference group (developed regions) in models 2 and 3. Additionally, the interaction terms with less developed regions are consistently negative and statistically significant across all specifications. These suggest that low-carbon investments contributed to smoothing short-term fluctuations in CO$_2$ emissions in these areas. The interaction terms with transition regions, however, are not statistically significant. For CO$_2$ intensity, no significant effects are observed for the reference group. However, the interaction terms with LDR are negative and statistically significant in models 5 and 6. For the 2014–2020 programming period, none of the estimated effects are statistically significant for either the reference group or interaction terms.
\begin{table}[h!]
\begin{center}
\scriptsize 
\caption{Summary of fixed-effects estimation results on emission cycles by gas type}\label{tb:model_cycle_types}\vspace{3pt}
\resizebox{0.7\textwidth}{!}{\renewcommand{\arraystretch}{1.1}
\begin{tabular}{lcccccc} 
\toprule 
 & \multicolumn{6}{c}{\textit{Dependent variables by gas type}} \\ 
\cmidrule{2-7} 
 & \multicolumn{3}{c}{CO$_2$ per capita} & \multicolumn{3}{c}{CO$_2$ intensity} \\ 
\cmidrule(lr){2-4}
\cmidrule(lr){5-7}
Model & (1) & (2) & (3) & (4) & (5) & (6)\\ 
\midrule
Low Carbon Exp. pc$^{07-13}_{t-5}$  & -0.010 & -0.066* & -0.081** & 0.034 & -0.007 & -0.018 \\
   & (0.029) & (0.037) & (0.034) & (0.047) & (0.038) & (0.035) \\
Low Carbon Exp. pc$^{14-20}_{t-5}$  & -0.174 & -0.224 & -0.237 & -0.137 & -0.203 & -0.217 \\
   & (0.273) & (0.293) & (0.258) & (0.291) & (0.291) & (0.273) \\
Low Carbon Exp. pc$^{07-13}_{t-5}$$\times$TER  & -0.206 & -0.163 & -0.166 & -0.148 & 0.015 & 0.008 \\
   & (0.165) & (0.145) & (0.143) & (0.197) & (0.113) & (0.118) \\
Low Carbon Exp. pc$^{14-20}_{t-5}$$\times$TER  & 0.325 & 0.090 & 0.127 & -0.113 & -0.478 & -0.459 \\
   & (0.880) & (0.865) & (0.823) & (0.948) & (0.844) & (0.841) \\
Low Carbon Exp. pc$^{07-13}_{t-5}$$\times$LDR  & -0.152** & -0.169*** & -0.199*** & -0.130* & -0.155*** & -0.173*** \\
   & (0.065) & (0.062) & (0.061) & (0.068) & (0.054) & (0.051) \\
Low Carbon Exp. pc$^{14-20}_{t-5}$$\times$LDR  & 0.295 & 0.350 & 0.410 & 0.181 & 0.278 & 0.316 \\
   & (0.285) & (0.302) & (0.274) & (0.298) & (0.291) & (0.275) \\
\midrule
 & \multicolumn{3}{c}{CH$_4$ per capita} & \multicolumn{3}{c}{CH$_4$ intensity} \\ 
\cmidrule(lr){2-4}
\cmidrule(lr){5-7}
Low Carbon Exp. pc$^{07-13}_{t-5}$  & -0.085*** & -0.073*** & -0.062*** & -0.041 & -0.014 & 0.000 \\
   & (0.025) & (0.023) & (0.022) & (0.027) & (0.028) & (0.031) \\
Low Carbon Exp. pc$^{14-20}_{t-5}$  & 0.252** & 0.148 & 0.145 & 0.289*** & 0.169* & 0.165* \\
   & (0.112) & (0.103) & (0.111) & (0.106) & (0.092) & (0.089) \\
Low Carbon Exp. pc$^{07-13}_{t-5}$$\times$TER  & -0.109 & -0.062 & -0.035 & -0.050 & 0.117** & 0.140*** \\
   & (0.094) & (0.065) & (0.058) & (0.127) & (0.047) & (0.053) \\
Low Carbon Exp. pc$^{14-20}_{t-5}$$\times$TER  & -0.527 & -0.614* & -0.667** & -0.965** & -1.182*** & -1.252*** \\
   & (0.354) & (0.337) & (0.334) & (0.411) & (0.304) & (0.337) \\
Low Carbon Exp. pc$^{07-13}_{t-5}$$\times$LDR  & 0.019 & 0.000 & 0.010 & 0.040 & 0.014 & 0.036 \\
   & (0.051) & (0.055) & (0.055) & (0.054) & (0.058) & (0.058) \\
Low Carbon Exp. pc$^{14-20}_{t-5}$$\times$LDR  & -0.256 & 0.037 & -0.002 & -0.369** & -0.035 & -0.097 \\
   & (0.172) & (0.173) & (0.184) & (0.170) & (0.178) & (0.177) \\
\midrule
 & \multicolumn{3}{c}{N$_2$O per capita} & \multicolumn{3}{c}{N$_2$O intensity} \\ 
\cmidrule(lr){2-4}
\cmidrule(lr){5-7}
Low Carbon Exp. pc$^{07-13}_{t-5}$  & -0.009 & -0.022 & -0.028 & 0.036 & 0.038 & 0.035 \\
   & (0.021) & (0.029) & (0.031) & (0.026) & (0.025) & (0.028) \\
Low Carbon Exp. pc$^{14-20}_{t-5}$  & 0.093 & 0.066 & 0.048 & 0.137 & 0.093 & 0.074 \\
   & (0.150) & (0.149) & (0.127) & (0.166) & (0.146) & (0.142) \\
Low Carbon Exp. pc$^{07-13}_{t-5}$$\times$TER  & -0.098** & -0.159*** & -0.163*** & -0.040 & 0.020 & 0.011 \\
   & (0.044) & (0.056) & (0.057) & (0.069) & (0.039) & (0.046) \\
Low Carbon Exp. pc$^{14-20}_{t-5}$$\times$TER  & 0.050 & 0.015 & 0.017 & -0.381 & -0.549*** & -0.563*** \\
   & (0.259) & (0.268) & (0.253) & (0.245) & (0.198) & (0.212) \\
Low Carbon Exp. pc$^{07-13}_{t-5}$$\times$LDR  & 0.113* & 0.129* & 0.121* & 0.134** & 0.143* & 0.147** \\
   & (0.063) & (0.073) & (0.072) & (0.064) & (0.073) & (0.073) \\
Low Carbon Exp. pc$^{14-20}_{t-5}$$\times$LDR  & -0.167 & -0.083 & -0.068 & -0.284* & -0.159 & -0.166 \\
   & (0.153) & (0.162) & (0.148) & (0.165) & (0.158) & (0.156) \\
\bottomrule
\multicolumn{7}{p{13cm}}{\textit{Notes:} Estimated coefficients of control variables are omitted. Model specifications correspond to those in Table~\ref{tb:model_trend}. Robust standard errors clustered at the NUTS 2 level are reported in parentheses. Significance levels: $^{*}$p$<$0.1; $^{**}$p$<$0.05; $^{***}$p$<$0.01.}\\
\end{tabular}}
\end{center}
\end{table}

Turning to the cycle component of CH$_4$ emissions (middle panel of Table~\ref{tb:model_cycle_types}), the 2007–2013 programme is associated with consistent and statistically significant reductions in per capita emissions for the reference group across all model specifications. In contrast, interaction terms with both TER and LDR are not statistically significant. In the case of the 2014–2020 programme, we observe a positive and statistically significant effect for the reference group in model 1 (CH$_4$ per capita), and across all models for CH$_4$ intensity (columns 4–6). The interaction terms with TER are negative and statistically significant in all models, except model 1. Interaction terms with LDR are not significant in most models, except CH$_4$ intensity in column 4, where a negative effect is detected.

Regarding the cycle component of N$_2$O emissions (lower panel of Table~\ref{tb:model_cycle_types}), the 2007–2013 programme shows no significant effects for the reference group in terms of per capita emissions. However, all estimated interaction terms with TER are negative and statistically significant, while those with LDR are positive and significant (columns 1-3). For N$_2$O intensity, the reference group shows no statistically significant response. However, the interaction terms with LDR are positive and statistically significant in models 4–6. For the 2014–2020 programming period, interaction terms with TER are again negative and statistically significant in models 5 and 6, and for LDR in model 1.

Therefore, the gas-specific results further support H1, showing that the effects of low-carbon expenditures differ across levels of regional development and are also highly heterogeneous across emission types. At the same time, the evidence also appears to support H2, since in less-developed regions, the low-carbon expenditures are positively associated with CO$_2$, CH$_4$, and N$_2$O trends, whereas in transition and developed regions, a negative relationship appears more common, although in many specifications this result is not statistically significant.

\section{Discussion}
\label{sec:discussion}

The results from the baseline fixed effects model reveal heterogeneous impacts of low-carbon investments on both emission trends and cycles. Generally, these investments do not yield statistically significant outcomes in more developed regions. However, in less developed areas, low-carbon investments appear paradoxically associated with increases in the long-term trend of per capita emissions, and, in some specifications, also in emissions intensity. This suggests that, in these regions, investments stimulate economic growth that remains dependent on carbon-intensive technologies. In contrast, when examining short-term fluctuations, the results suggest a more positive role for these investments in reducing emissions volatility, especially in less developed regions, indicating that the funds may help stabilise emissions in the short run.

When disaggregating the results by GHG type (CO$_2$, CH$_4$, and N$_2$O), a more nuanced picture emerges. For trends, the increase in per capita emissions observed in less developed regions is primarily driven by CO$_2$, the gas most closely linked to industrial activity and energy production. In contrast, the results for CH$_4$ and N$_2$O show mixed patterns, with adverse effects in some less developed regions and mitigating effects in transition areas. Regarding emission cycles, low-carbon investments show more consistently positive effects, particularly in reducing volatility for CO$_2$ in less developed regions and for CH$_4$ and N$_2$O in transition regions.

However, these results should be interpreted in light of the characteristics of the emission sources and policy contexts associated with each gas. Unlike CO$_2$, which is closely tied to fossil fuel combustion and therefore directly targeted by ESI-funded low-carbon investments, CH$_4$ and N$_2$O emissions are largely linked to agricultural practices and specific industrial processes \citep{scheffler2024eu}. Their responsiveness to low-carbon funding may thus be more limited or dependent on complementary policies. In particular, the Common Agricultural Policy (CAP), which is funded separately from the ESI funds, has distinct objectives that may create overlaps with climate-related goals \citep{european2021common}.

More broadly, the low-carbon economy fund from the ESI funds is one among several instruments in the EU's climate policy architecture. Notably, the EU ETS has played a more central role in recent decades \citep{bayer2020european}. Nevertheless, ESI funding remains important, as it aims to address regional disparities and supports place-based transitions.

Overall, the results are consistent with Hypothesis 1, indicating that low-carbon expenditures have heterogeneous impacts across different levels of regional development. Moreover, while we find consistent evidence that low-carbon economy funding does not lead to long-term reductions in emission indicators in more developed regions, the adverse effects observed in less developed areas are particularly concerning, given that these regions are the primary targets of ESI fund support. This pattern aligns with Hypothesis 2, which states that the effectiveness of such expenditures is conditional on regional development. Nevertheless, when examining the cyclical component of the emissions time series, we find that low-carbon investments may play a stabilising role by dampening short-term emissions volatility.

It is important to note that most statistically significant effects are concentrated in the 2007-2013 programming period. The lack of robust results for the 2014-2020 period may be due to data limitations. Given that the environmental impacts of investments often take time to materialise, the imprecision likely reflects the short post-investment observation window.

Moreover, in models using emissions intensity as the dependent variable, results are more frequently statistically not significant. Although this did not invalidate the central hypothesis of the study, it raises policy concerns. Specifically, the limited effects on emissions intensity, even in more developed areas, clearly hint that funding is not effectively supporting the decoupling of emissions from economic growth.

These findings call for a reassessment of EU climate investment strategies. On the one hand, the failure to curb long-term emissions in less developed areas raises questions about allocative efficiency and the risk of reinforcing carbon lock-in \citep{unruh2000understanding}. On the other hand, mechanisms behind the heterogeneous impacts---such as institutional capacity gaps, infrastructure rigidity, or lack of policy complementarities---remain underexplored and merit further attention \citep{rodriguez2015quality, charron2014regional}. In this respect, \cite{iammarino2019regional} emphasise the need for tailored policies that are sensitive to the structural prospects of different regions, aligning with the \textit{Barca report}'s advocacy for place-based strategies \citep{barca2009agenda}. 

In addition, while the observed short-term stabilisation of emissions is encouraging, it may reflect counter-cyclical spending effects rather than deeper structural transformations, thereby questioning the long-term decarbonisation potential of these funds.

Finally, although beyond this paper's scope, some of the emissions increases in less developed regions may reflect industrial relocation rather than policy failure. From a broader EU or global perspective, such shifts may be environmentally neutral. These trade-offs also underline the need for more targeted, region-specific strategies. Future research should investigate the mechanisms behind regional disparities and identify the complementary policies needed to align the ESI spending with long-term climate and development objectives.

It is worth noting that the identification strategy, based on long lags and fixed effects, is aimed to mitigate concerns of reverse causality; however, it may not fully address endogeneity. Instrumental variable estimations, using regional structural and institutional variables as instruments, resulted in weak instruments. Therefore, future work should seek to strengthen causal inference, for instance by exploiting exogenous quasi-natural experiments (e.g., changes in eligibility thresholds or policy allocation rules).

\section{Conclusions}
\label{sec:conclusions}

This paper examines the climate impact of one of the European Union's key instruments for regional development: the ``Low-Carbon Economy'' funds allocated through the European Structural and Investment funds during the 2007-2013 and 2014-2020 programming periods. Focusing on per capita emissions and emissions intensity---both decomposed into trend and cyclical components---we estimate the structural (long-run) and cyclical (short-run) effects of low-carbon expenditures across regions.

Our findings point to a substantial degree of heterogeneity. In more developed and transition regions, low-carbon funding shows no consistent long-term effect in reducing emissions, whether measured per capita or relative to GDP. However, such investments are associated with significant increases in long-term emissions trends in less developed regions, despite being the primary beneficiaries of these funds. These results suggest that, in lagging regions, climate-related investments may stimulate economic activity without achieving the intended decoupling from carbon-intensive processes. These findings are consistent across specifications with different sets of controls.

When disaggregating emissions by gas type (CO$_2$, CH$_4$, and N$_2$O), results show further heterogeneity. In less developed regions, increases in emissions are mainly associated with CO$_2$, whereas the effects on methane and nitrous oxide are less robust and vary across regions. Regarding short-run dynamics, we find some evidence that low-carbon investments contributed to reducing emissions volatility, especially related to expenditures of the 2007-2013 programme.

Overall, the empirical results raise questions about how effectively the EU Cohesion Policy supports the broader decarbonisation agenda. While our findings point to some limitations in the environmental impact of low-carbon investments, they also highlight the regional challenges in turning funding into measurable emission reductions. Structural conditions, institutional capacity, and how policies are implemented matter. However, other elements, such as the types of projects financed, their interaction with other policy initiatives, and the time it takes for their effects to materialise, likely play a role as well.

Our findings suggest that improving the climate effectiveness of the Cohesion Policy will require a more substantial alignment with regional contexts and development trajectories. Given the heterogeneous impacts observed across regions, it is essential to move beyond uniform approaches.  Less developed regions deserve particular attention. Although they are not currently the highest emitters on a per capita basis, they face the risk of becoming major sources of emissions as they develop. We claim that aligning investment strategies with regional profiles can enhance the effectiveness of emissions reduction efforts.

\section*{Acknoledgment}
The authors acknowledge the financial support of the DECIPHER (Decision-making framework and processes for holistic evaluation of environmental and climate policies) project, funded by the Horizon Europe: Research and Innovation Programme under grant agreement No. 101056898.

\clearpage
\newpage
\bibliographystyle{chicago}
\bibliography{biblio_funds}

\clearpage
\newpage

\appendix
\section*{Appendix}

\setcounter{section}{0}\renewcommand{\thesection}{A\arabic{section}}
\setcounter{table}{0}\renewcommand{\thetable}{A\arabic{table}}
\setcounter{figure}{0}\renewcommand{\thefigure}{A\arabic{figure}}
\setcounter{equation}{0}\renewcommand{\theequation}{A\arabic{equation}}

\section{Low-carbon economy definitions by ESI programmes}
\label{app:definitions}

\begin{table}[h!]
\begin{center}
\scriptsize 
\caption{Definitions of low-carbon economy expenditures.}
\label{tb:low_carbon_def}\vspace{3pt}
\resizebox{0.7\textwidth}{!}{\renewcommand{\arraystretch}{1.1}
\begin{tabular}{c >{\raggedright\arraybackslash}p{8cm}}
\toprule 
\multicolumn{2}{c}{\textit{2014-2020 funding program}} \\ 
\midrule
Intervention field code & Intervention field description \\
\midrule
009 & Renewable energy: wind \\
010 & Renewable energy: solar \\
011 & Renewable energy: biomass \\
012 & Other renewable energy (including hydroelectric, geothermal and marine energy) and renewable energy integration (including storage, power to gas and renewable hydrogen infrastructure) \\
013 & Energy efficiency renovation of public infrastructure, demonstration projects and supporting measures \\
014 & Energy efficiency renovation of existing housing stock, demonstration projects and supporting measures \\
015 & Intelligent Energy Distribution Systems at medium and low voltage levels (including smart grids and ICT systems) \\
016 & High efficiency co-generation and district heating \\
023 & Environmental measures aimed at reducing and / or avoiding greenhouse gas emissions (including treatment and storage of methane gas and composting) \\
043 & Clean urban transport infrastructure and promotion (including equipment and rolling stock) \\
068 & Energy efficiency and demonstration projects in SMEs and supporting measures \\
070 & Promotion of energy efficiency in large enterprises \\
090 & Cycle tracks and footpaths \\
\midrule
\multicolumn{2}{c}{\textit{2007-2013 funding program}} \\ 
\midrule
Priority code & Priority description \\
\midrule
06 & Assistance to small and medium enterprises (SMEs) for the promotion of environmentally-friendly products and production processes (...) \\
24 & Cycle tracks \\
28 & Intelligent transport systems \\
39 & Renewable energy: wind \\
40 & Renewable energy: solar \\
41 & Renewable energy: biomass \\
42 & Renewable energy: hydroelectric, geothermal and other \\
43 & Energy efficiency, co-generation, energy management \\
25-52 & Promotion of clean urban transport \\
\bottomrule
\end{tabular}}
\end{center}
\end{table}

\clearpage
\newpage
\section{Estimation results modelled expenditures}
\label{app:est_modelled}

The cohesion data platform presents estimations of the expenditure by NUTS 2, which are aggregated for the most significant types of structural funds only. Therefore, to address our need for yearly data, we have opted for two approaches: the one presented in the main text, which assumes that expenditures are distributed evenly over time (dividing the reported expenditures over the programming period), and the another one, presented here, which assumes they follow the same pattern as the total aggregated expenditure reported by the cohesion data platform.\footnote{Data available at: \url{https://cohesiondata.ec.europa.eu/Other/Historic-EU-payments-regionalised-and-modelled/tc55-7ysv/about_data}.} Both strategies yielded qualitatively similar conclusions, as a robustness check the results of the second strategy are shown in Tables~\ref{tb:model_trend_tm} and \ref{tb:model_cycle_tm}.
\begin{table}[h!]
\begin{center}
\scriptsize 
\caption{Fixed effect estimation results on GHG emissions trends}\label{tb:model_trend_tm}\vspace{3pt}
\resizebox{0.8\textwidth}{!}{\renewcommand{\arraystretch}{1.1}
\begin{tabular}{lcccccc} 
\toprule 
 & \multicolumn{6}{c}{\textit{Dependent variable:}} \\ 
\cmidrule{2-7} 
 & \multicolumn{3}{c}{GHGEpc} & \multicolumn{3}{c}{GHGEi} \\ 
\cmidrule(lr){2-4}
\cmidrule(lr){5-7}
& (1) & (2) & (3) & (4) & (5) & (6)\\ 
\midrule
Low Carbon Exp. pc$^{07-13}_{t-5}$  & 0.057 & -0.034 & 0.111 & -0.530*** & -0.141 & -0.070 \\
   & (0.231) & (0.190) & (0.134) & (0.182) & (0.199) & (0.164) \\
Low Carbon Exp. pc$^{14-20}_{t-5}$  & -0.282 & -0.622 & 0.603 & -2.513 & -1.596 & -0.850 \\
   & (1.654) & (2.160) & (1.928) & (2.621) & (2.037) & (1.896) \\
Low Carbon Exp. pc$^{07-13}_{t-5}$$\times$TER  & -1.500 & -0.665 & -0.955 & 0.064 & -0.571 & -0.646 \\
   & (0.951) & (1.117) & (1.213) & (0.818) & (0.748) & (0.809) \\
Low Carbon Exp. pc$^{14-20}_{t-5}$$\times$TER  & -3.332 & -1.695 & -4.412 & 1.612 & -1.917 & -3.485 \\
   & (4.645) & (4.536) & (4.570) & (4.472) & (4.400) & (4.136) \\
Low Carbon Exp. pc$^{07-13}_{t-5}$$\times$LDR  & 0.997*** & 0.879*** & 0.701*** & 0.548* & 0.544* & 0.447 \\
   & (0.299) & (0.268) & (0.247) & (0.298) & (0.328) & (0.312) \\
Low Carbon Exp. pc$^{14-20}_{t-5}$$\times$LDR  & 2.964* & 2.601 & 1.533 & 1.442 & 1.766 & 1.011 \\
   & (1.673) & (1.989) & (1.816) & (2.480) & (1.894) & (1.795) \\
GDP$_{t-5}$  &  & 0.240*** &  &  & -0.467*** &  \\
   &  & (0.065) &  &  & (0.102) &  \\
EPS$_{t-5}$  &  & 0.001 & 0.002 &  & 0.003 & 0.002 \\
   &  & (0.007) & (0.006) &  & (0.007) & (0.007) \\
Political Ideology$_{t-5}$  &  & 0.016 & 0.008 &  & -0.010 & -0.008 \\
   &  & (0.010) & (0.009) &  & (0.016) & (0.014) \\
GVA$_{A,t-5}$  &  &  & 0.067** &  &  & 0.038 \\
   &  &  & (0.033) &  &  & (0.036) \\
GVA$_{B-E,t-5}$  &  &  & 0.246*** &  &  & -0.038 \\
   &  &  & (0.047) &  &  & (0.075) \\
GVA$_{F,t-5}$  &  &  & -0.136*** &  &  & -0.155*** \\
   &  &  & (0.041) &  &  & (0.057) \\
GVA$_{G-J,t-5}$  &  &  & 0.129 &  &  & -0.205 \\
   &  &  & (0.090) &  &  & (0.131) \\
GVA$_{K-N,t-5}$  &  &  & -0.366*** &  &  & -0.392*** \\
   &  &  & (0.084) &  &  & (0.099) \\
GVA$_{O-U,t-5}$  &  &  & 0.167 &  &  & 0.302** \\
 &  &  & (0.109) &  &  & (0.134) \\
\hline
Observations & 3,872 & 3,408 & 3,408 & 3,872 & 3,408 & 3,408 \\
\bottomrule
\multicolumn{7}{p{14cm}}{\textit{Notes:} NUTS 2 cluster robust standard errors are in parentheses. Significance level: $^{*}$p$<$0.1; $^{**}$p$<$0.05; $^{***}$p$<$0.01}\\
\end{tabular}}
\end{center}
\end{table}

\begin{table}[h!]
\begin{center}
\scriptsize 
\caption{Fixed effect estimation results on GHG emissions cycles}\label{tb:model_cycle_tm}\vspace{3pt}
\resizebox{0.8\textwidth}{!}{\renewcommand{\arraystretch}{1.1}
\begin{tabular}{lcccccc} 
\toprule 
 & \multicolumn{6}{c}{\textit{Dependent variable:}} \\ 
\cmidrule{2-7} 
 & \multicolumn{3}{c}{GHGEpc} & \multicolumn{3}{c}{GHGEi} \\ 
\cmidrule(lr){2-4}
\cmidrule(lr){5-7}
& (1) & (2) & (3) & (4) & (5) & (6)\\ 
\midrule
Low Carbon Exp. pc$^{07-13}_{t-5}$  & 0.034 & -0.021 & -0.031 & 0.056* & 0.038 & 0.026 \\
   & (0.024) & (0.031) & (0.031) & (0.033) & (0.025) & (0.026) \\
Low Carbon Exp. pc$^{14-20}_{t-5}$  & -0.671 & -0.713 & -0.779 & -0.145 & -0.569 & -0.602 \\
   & (0.722) & (0.804) & (0.722) & (0.740) & (0.644) & (0.615) \\
Low Carbon Exp. pc$^{07-13}_{t-5}$$\times$TER  & -0.290** & -0.270*** & -0.281*** & -0.224 & -0.091 & -0.095 \\
   & (0.117) & (0.081) & (0.083) & (0.152) & (0.074) & (0.081) \\
Low Carbon Exp. pc$^{14-20}_{t-5}$$\times$TER  & 1.702 & 1.176 & 1.261 & -0.321 & -0.915 & -0.914 \\
   & (1.997) & (1.990) & (1.900) & (2.445) & (2.246) & (2.276) \\
Low Carbon Exp. pc$^{07-13}_{t-5}$$\times$LDR  & -0.143** & -0.141** & -0.157*** & -0.127** & -0.130*** & -0.138*** \\
   & (0.058) & (0.056) & (0.056) & (0.053) & (0.046) & (0.046) \\
Low Carbon Exp. pc$^{14-20}_{t-5}$$\times$LDR  & 0.805 & 0.985 & 1.082 & 0.323 & 0.793 & 0.844 \\
   & (0.757) & (0.809) & (0.745) & (0.768) & (0.668) & (0.642) \\
GDP$_{t-5}$  &  & -0.251*** &  &  & 0.080** &  \\
   &  & (0.037) &  &  & (0.035) &  \\
EPS$_{t-5}$  &  & 0.009*** & 0.010*** &  & 0.011*** & 0.012*** \\
   &  & (0.003) & (0.003) &  & (0.003) & (0.003) \\
Political Ideology$_{t-5}$  &  & 0.004 & 0.002 &  & 0.002 & 0.002 \\
   &  & (0.003) & (0.003) &  & (0.003) & (0.003) \\
GVA$_{A,t-5}$  &  &  & -0.011 &  &  & -0.011 \\
   &  &  & (0.009) &  &  & (0.007) \\
GVA$_{B-E,t-5}$  &  &  & -0.011 &  &  & -0.004 \\
   &  &  & (0.018) &  &  & (0.018) \\
GVA$_{F,t-5}$  &  &  & -0.044*** &  &  & 0.017 \\
   &  &  & (0.014) &  &  & (0.017) \\
GVA$_{G-J,t-5}$  &  &  & 0.028 &  &  & 0.078** \\
   &  &  & (0.033) &  &  & (0.032) \\
GVA$_{K-N,t-5}$  &  &  & -0.049 &  &  & 0.024 \\
   &  &  & (0.040) &  &  & (0.038) \\
GVA$_{O-U,t-5}$  &  &  & -0.197*** &  &  & -0.057 \\
 &  &  & (0.042) &  &  & (0.042) \\
\hline
Observations & 3,872 & 3,408 & 3,408 & 3,872 & 3,408 & 3,408 \\
 \bottomrule
\multicolumn{7}{p{14cm}}{\textit{Notes:} NUTS 2 cluster robust standard errors are in parentheses. Significance level: $^{*}$p$<$0.1; $^{**}$p$<$0.05; $^{***}$p$<$0.01}\\
\end{tabular}}
\end{center}
\end{table}

\clearpage
\newpage
\section{Instrumental variables (IV/2SLS)}
\label{app:iv_results}

An instrumental variables approach can be used to address potential endogeneity concerns related to low-carbon investments as a complement to our baseline fixed-effects specification. While the use of lagged explanatory variables in the baseline models helps mitigate simultaneity bias, endogeneity may persist due to omitted variables. For instance, more proactive regions might both attract higher low-carbon funding and implement complementary environmental measures not fully captured by our controls, potentially biasing the estimated coefficients.

However, one of the main challenges in identifying valid instruments for low-carbon investments is that, while several structural variables can plausibly explain the allocation of such funds without being directly correlated with emissions, the delayed nature of their effects substantially weakens the instruments' explanatory power. As a result, the fitted values of investment exhibit limited variation, which reduces the likelihood of detecting statistically significant effects in the second stage. We present these results below.

We first identified a set of potential instruments for per capita low-carbon expenditures based on regional structural indicators from the ARDECO database, including labour productivity per employee (log), consumption of fixed capital (log), hours worked per capita, and compensation of employees (log). Then, we employed two-stage least squares estimations (2SLS), including year, regional, and eligibility fixed effects, clustering standard errors at the regional level, and controlling for lagged sectoral value added and structural indicators.

After testing various specifications, we found that, according to the underidentification test, these variables performed reasonably well when considered either individually or in combination. Subsequently, we observed that using three- to five-period lags yielded comparatively better first-stage results. It is not surprising that lagged versions of these variables perform well, as pre-existing structural conditions are likely to influence the allocation of funds. However, these lags are added to the five-period lag already applied to the dependent variable, weakening their potential as strong instruments. 

In Table~\ref{tb:iv_app}, we report the results for two specifications: the trend of GHG emissions per capita (third column) and the trend of GHG emissions intensity (fourth column), along with their corresponding first-stage regressions (first two columns). We present these estimations as an example to illustrate the limitations of our empirical approach. In these specifications, the set of instruments is reasonably correlated with the endogenous regressor, as indicated by the Kleibergen-Paap rk LM statistic (greater than 10, with Chi-sq. $p$-value $<$ 0.05), and there is no evidence against their validity according to the Hansen J test ($p$-value $>$ 0.1). However, the instruments consistently fail to pass the weak identification tests: the Kleibergen-Paap rk Wald $F$ statistic remains well below conventional thresholds ($F > 10$). Because this statistic is robust to heteroskedasticity and clustering, it provides the appropriate benchmark in our panel setting. 

After implementing different specifications to exhaustively explore the possibility of addressing potential endogeneity concerns, we therefore conclude that the IV estimates cannot provide reliable evidence and must be interpreted with caution.
\begin{table}[h!]
\begin{center}
\scriptsize 
\caption{IV/2SLS estimations}\label{tb:iv_app}\vspace{3pt}
\resizebox{0.9\textwidth}{!}{\renewcommand{\arraystretch}{1.1}
\begin{tabular}{l >{\centering\arraybackslash}m{2.5cm} >{\centering\arraybackslash}m{2.5cm} >{\centering\arraybackslash}m{2.5cm} >{\centering\arraybackslash}m{2.5cm}} 
\toprule 
 & \multicolumn{2}{c}{First stage} & \multicolumn{2}{c}{Second stage} \\ 
\cmidrule(lr){2-3}
\cmidrule(lr){4-5}
 & \multicolumn{4}{c}{\textit{Dependent variable:}} \\ 
\cmidrule{2-5} 
& Low Carbon Exp. pc$^{07-13}_{t-5}$ & Low Carbon Exp. pc$^{14-20}_{t-5}$ & GHGE per capita & GHGE intensity\\ 
\midrule
Low Carbon Exp. pc$^{07-13}_{t-5}$  &  &  & -0.604 & -0.101 \\
   &  &  & (0.588) & (0.615) \\
Low Carbon Exp. pc$^{14-20}_{t-5}$  &  &  & 3.071 & -7.180 \\
   &  &  & (2.064) & (4.764) \\
Low Carbon Exp. pc$^{07-13}_{t-5}$$\times$TER  &  &  & 1.558* & 2.676** \\
   &  &  & (0.888) & (1.272) \\
Low Carbon Exp. pc$^{14-20}_{t-5}$$\times$TER  &  &  & -3.332 & -3.195 \\
   &  &  & (4.136) & (5.033) \\
Low Carbon Exp. pc$^{07-13}_{t-5}$$\times$LDR  &  &  & 0.535 & 0.446 \\
   &  &  & (0.693) & (0.726) \\
Low Carbon Exp. pc$^{14-20}_{t-5}$$\times$LDR  &  &  & 2.078 & 7.318 \\
   &  &  & (2.252) & (4.569) \\
EPS$_{t-5}$  & 0.018*** & 0.003** & -0.004 & -0.015 \\
   & (0.004) & (0.001) & (0.010) & (0.011) \\
Political Ideology$_{t-5}$  & 0.016*** & 0.004*** & -0.007 & -0.016 \\
 & (0.006) & (0.001) & (0.014) & (0.017) \\
GVA$_{A,t-6}$  & -0.005 & -0.001 & 0.069* & 0.033 \\
   & (0.020) & (0.003) & (0.035) & (0.039) \\
GVA$_{B-E,t-6}$  & 0.023 & 0.010* & 0.142** & -0.078 \\
   & (0.028) & (0.005) & (0.070) & (0.092) \\
GVA$_{F,t-6}$  & -0.017 & -0.001 & -0.134*** & -0.126 \\
   & (0.018) & (0.007) & (0.051) & (0.079) \\
GVA$_{G-J,t-6}$  & -0.084 & -0.031*** & 0.376*** & -0.058 \\
   & (0.059) & (0.011) & (0.134) & (0.167) \\
GVA$_{K-N,t-6}$  & 0.051 & 0.057*** & -0.444*** & -0.257** \\
   & (0.035) & (0.013) & (0.104) & (0.102) \\
GVA$_{O-U,t-6}$  & -0.128** & 0.015 & 0.112 & 0.233* \\
 & (0.052) & (0.014) & (0.109) & (0.137) \\
Labour productivity ppe$_{t-8}$  & 0.238*** & 0.029* &  &  \\
 & (0.063) & (0.017) &  &  \\
Labour productivity ppe$_{t-8}$$\times$TER  & -0.217*** & -0.024 &  &  \\
 & (0.066) & (0.021) &  &  \\
Labour productivity ppe$_{t-8}$$\times$LDR & -0.149** & -0.027 &  &  \\
 & (0.060) & (0.017) &  &  \\
Consumption of fixed capital$_{t-8}$  & -0.047 & -0.059*** &  &  \\
 & (0.034) & (0.013) &  &  \\
Consumption of fixed capital$_{t-8}$$\times$TER  & 0.174*** & 0.046*** &  &  \\
 & (0.030) & (0.012) &  &  \\
Consumption of fixed capital$_{t-8}$$\times$LDR & 0.091*** & 0.009 &  &  \\
 & (0.025) & (0.010) &  &  \\
Hours worked per capita$_{t-8}$  & 0.000 & 0.000 &  &  \\
 & (0.000) & (0.000) &  &  \\
Hours worked per capita$_{t-8}$$\times$TER  & -0.000 & 0.000 &  &  \\
 & (0.000) & (0.000) &  &  \\
Hours worked per capita$_{t-8}$$\times$LDR & 0.000 & 0.000 &  &  \\
 & (0.000) & (0.000) &  &  \\
Compensation of Employees$_{t-8}$  & 0.125*** & 0.011 &  &  \\
 & (0.037) & (0.016) &  &  \\
Compensation of Employees$_{t-8}$$\times$TER  & -0.047 & -0.013 &  &  \\
 & (0.058) & (0.024) &  &  \\
Compensation of Employees$_{t-8}$$\times$LDR & -0.156*** & -0.005 &  &  \\
 & (0.052) & (0.021) &  &  \\
 & -1.384** & -0.315** &  &  \\
 & (0.633) & (0.157) &  &  \\
\midrule
Observations & 3,408 & 3,408 & 3,408 & 3,408 \\
\midrule
Kleibergen-Paap rk LM (p-value) &  &  & 26.844 (0.000) & 26.844 (0.000) \\
Cragg–Donald Wald F &  &  & 12.957 & 12.957 \\
Kleibergen-Paap rk Wald F &  &  & 2.676 & 2.676 \\
Hansen J (p-value) &  &  & 7.542 (0.274) & 9.196 (0.163) \\
\bottomrule
\multicolumn{5}{p{17cm}}{\textit{Notes:} Robust standard errors clustered at the NUTS 2 level are reported in parentheses. Significance level: $^{*}$p$<$0.1; $^{**}$p$<$0.05; $^{***}$p$<$0.01.}\\
\end{tabular}}
\end{center}
\end{table}

\end{document}